\documentclass[sigconf,screen]{acmart}

\usepackage{subfig}
\usepackage{stfloats}
\usepackage{enumitem}

\usepackage{url}
\usepackage[normalem]{ulem}

\usepackage{amsmath,amssymb,amsfonts}
\usepackage{algorithmic}
\usepackage{multirow}
\usepackage{stfloats}

\usepackage{amsthm}
\usepackage{braket}
\usepackage{tcolorbox}

\usepackage{etoolbox}
\AtBeginEnvironment{tabular}{\footnotesize}

\definecolor{aliceblue}{rgb}{0.94, 0.97, 1.0}
\usepackage{xcolor}
\usepackage{tcolorbox}
\tcbset{width=1\columnwidth,boxrule=1pt,colback=aliceblue,arc=1pt,left=2pt,right=2pt,boxsep=0.5pt}

\usepackage[flushleft]{threeparttable}

\newcommand{\ignore}[1]{}

\usepackage{url}
\AtBeginDocument{%
  }

\setcopyright{none}
\copyrightyear{2023}
\acmYear{2023}
\acmDOI{XXXXXXX.XXXXXXX}

\acmConference[ASPLOS '23]{Proceedings of the 28th ACM International Conference on Architectural Support for Programming Languages and Operating Systems}{March 25 -- March 29, 2023}{Vancouver, Canada }

\begin{document}

\title[\empty]{
  FrozenQubits:   
  Boosting Fidelity of QAOA
  \\by Skipping Hotspot Nodes
}

\settopmatter{authorsperrow=2}
\author{Ramin Ayanzadeh$^*$}
\affiliation{%
  \institution{Georgia Institute of Technology}
  \city{Atlanta}
  \country{USA}
}

\author{Narges Alavisamani}
\affiliation{%
\institution{Georgia Institute of Technology}
  \city{Atlanta}
  \country{USA}
}

\author{Poulami Das}
\affiliation{%
\institution{Georgia Institute of Technology}
  \city{Atlanta}
  \country{USA}
}

\author{Moinuddin Qureshi}
\affiliation{%
\institution{Georgia Institute of Technology}
  \city{Atlanta}
  \country{USA\vspace{0.15in}}
}
\renewcommand{\shortauthors}{\empty}

\begin{abstract}
    Quantum Approximate Optimization Algorithm (QAOA) is one of the leading candidates for demonstrating the quantum advantage using near-term quantum computers. Unfortunately, high device error rates limit us from reliably running QAOA circuits for problems with more than a few qubits. 
    In QAOA, the problem graph is translated into a quantum circuit such that every edge corresponds to two 2-qubit CNOT operations in each layer of the circuit. As CNOTs are extremely error-prone, the fidelity of QAOA circuits is dictated by the number of edges in the problem graph. 
    
    We observe that majority of graphs corresponding to real-world applications follow the ``power-law`` distribution, where some hotspot nodes have significantly higher number of connections. 
    We leverage this insight and propose ``FrozenQubits`` that freezes the hotspot nodes or qubits and intelligently partitions the state-space of the given problem into several smaller sub-spaces which are then solved independently. 
    The corresponding QAOA sub-circuits are significantly less vulnerable to gate and decoherence errors due to the reduced number of CNOT operations in each sub-circuit. 
    Unlike prior circuit-cutting approaches, FrozenQubits does not require any exponentially complex post-processing step. 
    Our evaluations with 5,300 QAOA circuits on eight different quantum computers from IBM shows that FrozenQubits can improve the quality of solutions by 8.73x on average  (and by up to 57x), albeit utilizing 2x more quantum resources.  
    
\end{abstract}

\settopmatter{printfolios=true} 

\maketitle
\pagestyle{plain}

\section{Introduction}
Near-term quantum computers with few dozens of noisy qubits can already outperform supercomputers for certain tasks~\cite{arute2019quantum,villalonga2020establishing}. 
Soon, we expect these Noisy Intermediate Scale Quantum (NISQ)~\cite{preskillNISQ} devices to provide computational advantages for real-world applications such as estimating the ground state energy of molecules~\cite{nam2020ground,preskillNISQ} and discrete optimizations~\cite{farhi2014quantum}. 
Unfortunately, NISQ programs are vulnerable to errors due to noise and their fidelity is low. To leverage NISQ systems for practical problems, the application fidelity must be increased.

\begin{figure*}[thb]
	\vspace{0.2in}
	\centering
	\includegraphics[width=\textwidth]{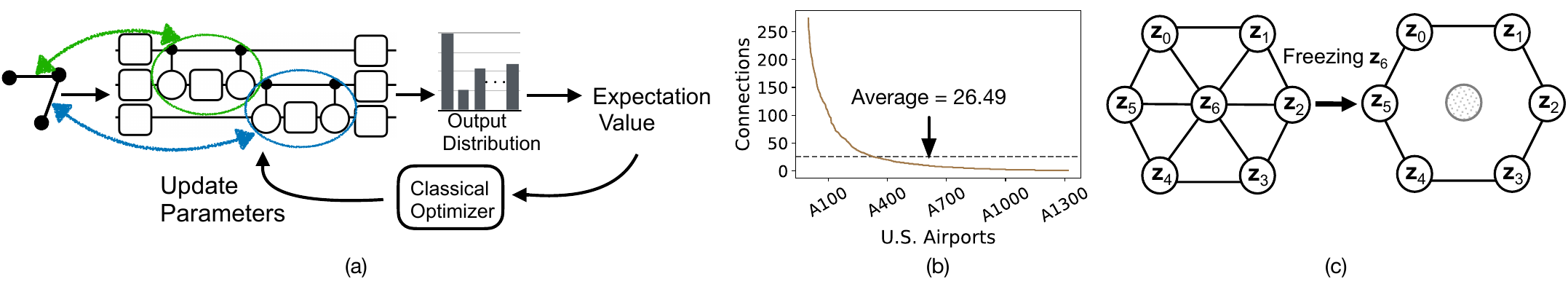}
    \caption{
		(a)~Problem solving using QAOA. 
		(b)~Node degree of the U.S. airport connections follows the power-law distribution where Hotspots or hubs have significantly higher connectivity than the average.
		(c)~Freezing $\mathbf{z}_6$ drops 6 edges.
	}
    \vspace{0.2in}
	\label{fig:intro_fig}
\end{figure*}

{\em Quantum approximate optimization algorithm} (QAOA)~\cite{basso2021quantum,farhi2014quantum} is regarded as a leading candidate for demonstration of quantum advantage on NISQ devices. 
It approximates the ground state or a configuration with the lowest energy value of a physical system, called \emph{Hamiltonian}. 
QAOA promises computational advantages for various industry-scale applications~\cite{choi2020quantum,cordier2021biology,dalyac2021qualifying,streif2021beating,vikstaal2020applying}, where the objective is to minimize or maximize a cost function.
Solving a problem using QAOA is a two-step process, as shown in Figure~\ref{fig:intro_fig}(a). 
First, the problem is mapped into a $p$-layer parametric quantum circuit with $2p$ parameters and executed for thousands of trials. 
Next, the expectation value of the output distribution is  used by a classical optimizer to adjust the parameters. 
The process is repeated until the optimal parameters of the circuit have been found. 
Unfortunately, noisy devices limit us from running QAOA problems of practical scale. 
For example, despite various optimizations for reducing the impact of hardware errors, we are unable to solve QAOA problems on 3-regular graphs with more than twenty-three qubits on state-of-the-art NISQ systems such as Google Sycamore~\cite{harrigan2021quantum}. 


The number of two-qubit CNOT operations in a QAOA circuit increases linearly with the number of edges in the problem graph, as each edge typically corresponds to two CNOTs, as shown in Figure~\ref{fig:intro_fig}(a). 
CNOT operations are highly error-prone and incur long latencies. 
For example, CNOTs have an average error-rate of 1\% on Google Sycamore~\cite{sycamoredatasheet} and take 400ns on average on IBM devices (10x slower than single-qubit gates). 
The problem compounds as most NISQ devices have limited connectivity and compilers are forced to introduce SWAPs that further increase the number of CNOTs required and depth of circuits. 
Post-compilation, the number of CNOTs scales super-polynomial with the number of edges in the problem graph. 
The high error-rates of CNOT operations and inability of qubits to retain information beyond a few micro-seconds on near-term systems 
cause the infeasability of practical QAOA problems~\cite{guerreschi2019qaoa} to become vanishingly small as a result of large CNOT count and circuit depth.

We observe that although most QAOA studies focus on 3-regular or fully-connected graphs, real-world problems follow {\em power-law  (or Pareto) distribution}~\cite{agler2016microbial,clauset2016colorado,gamermann2019comprehensive,goh2002classification,house2015testing,mislove2007measurement,pastor2015epidemic}. 
This means only a small number of nodes have much higher connectivity than others. 
For example, Figure~\ref{fig:intro_fig}(b) shows the number of connections for 1300 U.S. airports. 
We observe that some airports (known as hotspots or hubs) have a significantly larger number of connections or edges than others. 
The ten busiest airports have 10x connections compared to the average connectivity of all airports. 
From the QAOA perspective, these nodes with significantly higher connectivity than the average contribute to a greater number of CNOT operations in the parametric circuit than others. 
We leverage this insight and propose {\em FrozenQubits}  to improve the fidelity of QAOA problems. 




In FrozenQubits, we freeze some of the nodes with the highest degree of connectivity. 
This drops the edges connected to these nodes resulting in a circuit with  fewer CNOTs and lower depth. 
We explain our design using the example shown in Figure~\ref{fig:intro_fig}(c). 
Here, node $\mathbf{z}_6$ is connected to six other nodes and contributes to 12 CNOTs in the original QAOA circuit. 
Note that this is under the assumption that the device topology is such that the compiler does not need to introduce any SWAPs. 
On real NISQ machines with limited connectivity, however, the CNOT count and depth will be even higher. 
The probability of introducing SWAPs for a node increases with the number of connections. Therefore, for QAOA circuits, hotspot nodes with significantly higher connectivity typically tend to have larger SWAP overheads compared to nodes with lower connectivity. 
Our proposed solution freezes node $\mathbf{z}_6$ and creates two sub-circuits, resulting in 50\% reduction in the number of CNOTs in each sub-circuit. 
This also eliminates any SWAPs that would have been otherwise required for the CNOTs corresponding to the edges for node $\mathbf{z}_6$. 
The reduction in SWAP overheads will depend on the topology of the target device.

In FrozenQubits, we identify a hotspot node (which corresponds to qubits with the highest number of CNOTs), freeze it by substituting the value of each qubit with its two possibilities, -1 and +1. 
This partitions the state-space of the given problem into two smaller problems whose corresponding QAOA circuits are significantly less vulnerable to errors, and solve the resultant sub-circuits independently. 
In general, freezing ${m}$ qubits will result in $2^m$ circuits. Note that, usually problem graphs have only a few hotspots, as shown in Figure~\ref{fig:intro_fig}(b). Thus, a small value of $m$ is sufficient and our default design uses $m =1, 2$. 
The challenge with FrozenQubits is that we do not know which sub-circuit includes the global optimum and therefore, we need to run all $2^m$ sub-circuits to find the solution. 
However, we observe that majority of these sub-circuits are pairwise symmetric whereby flipping all bits of any point in one sub-space is a point in the other sub-space with an identical cost. 
We leverage this insight to reduce the complexity of FrozenQubits by only solving one of these pairwise symmetric sub-circuits and then inferring the result of the other sub-circuit. 
For example, freezing $\mathbf{z}_6$ in Figure~\ref{fig:intro_fig}(c) results in two smaller circuits, one for $\mathbf{z}_6=+1$ and one for $\mathbf{z}_6=-1,$ 
but we only evaluate one of them and postprocess its output distribution to evaluate the other circuit.

We note that FrozenQubits is significantly different than circuit cutting. 
For example, with CutQC~\cite{tang2021cutqc}, a circuit is decomposed into $2^{c}$ sub-circuits by cutting ${c}$ edges from the circuit graph and the sub-circuits are executed separately. The output distribution is obtained using tensor products. 
While this approach is promising, it is not applicable to practical QAOA circuits because: 
(1) for real-world graphs, ${c}$ must be greater than the number of hotspots, which is impractical as the number of sub-circuits and complexity of the post-processing step grow exponentially with ${c}$,
(2) partitioning multi-layer QAOA circuits via cutting wires is nontrivial, 
and (3) exponentially complex post-processing can bottleneck the tuning of circuit parameters. On the contrary, FrozenQubits does not require evaluation of all the sub-circuits or involve any post-processing step that incurs exponential complexity.
Similarly, edge cutting has been proposed for partitioning large graphs into smaller sub-graphs and running the sub-graphs on smaller quantum computers~\cite{li2021large}. 
However, edge-cutting power-law graphs is nontrivial as hotspot nodes appear in all sub-graphs that degrades the accuracy of the output distribution estimation.
Instead, we take an orthogonal approach in this paper to simplify the search space of QAOA.

Our evaluations with over 5,300 circuits on eight different real quantum computers from IBM show that FrozenQubits can obtain by 8.73x on average (and up to 57.14x) improvement in the quality of solutions for QAOA circuits compared to the baseline, albeit running 2x more circuits. 
FrozenQubits data and code are publicly available at 
\href{https://doi.org/10.5281/zenodo.7278397}{https://doi.org/10.5281/zenodo.7278397}.

\vspace{0.05in}
Overall, this paper makes the following contributions:
\begin{enumerate}
	
	\item  We propose \emph{FrozenQubits}, a novel quantum divide-and-conquer approach that leverages the insight about most real-world applications having power-law graphs to boost the fidelity of QAOA applications; 

	\item We observe that the search space of most QAOA problems have symmetric sub-spaces whereby flipping all bits of any point in one sub-space is a point in the other sub-space with an identical cost value; 

	\item We leverage the symmetricity of the search space of most QAOA problems  to subside the complexity of our design.
\end{enumerate}

\newpage
\section{Background}

\subsection{ Quantum Approximate Optimization Algorithm}
{\em Quantum approximate optimization algorithm {\em(QAOA})}~\cite{basso2021quantum,farhi2014quantum,farhi2014quantum_applied} is widely recognised as one of the leading candidates for demonstrating quantum advantage in the near-term. 
QAOA promises computational speed-up for optimization problems in various application domains. 
Solving real-world optimization problems using QAOA is a two-step process. 
First, representing the optimization problem in the form of Equation~\eqref{eqn:Ising}, which is known as \textbf{\textit{Ising Hamiltonian}}, 
where $\mathbf{z}_i \in \{ {-1,+1} \}$ and $\mathbf{h}_i, J_{ij}, \textit{offset} \in \mathbb{R}$ 
are given coefficients of the problem~\cite{ayanzadeh2021equal,ayanzadeh2020multi,ayanzadeh2022quantum,ayanzadeh2020ensemble,ayanzadeh2020reinforcement,ayanzadeh2019quantum,das2008colloquium,harrigan2021quantum,lucas2014ising,o2018nonnegative,rieffel2015case,sherrington1975solvable,sung2020using}.
\begin{equation}
	H_Z := C({\mathbf{z}}) = \sum_{i=0}^{N-1}{\mathbf{h}_i \mathbf{z}_i} +
		\sum_{i=0}^{N-1}{ \sum_{j=i+1}^{N-1}{J_{ij} \mathbf{z}_i \mathbf{z}_j} }
		+\text{offset}
	\label{eqn:Ising}
\end{equation}
Second, tuning QAOA circuit parameters to find the optimum value of Equation~\eqref{eqn:Ising}. 
QAOA promises speed-up for many applications by accelerating this step, which is computationally expensive on conventional computers~\cite{farhi2014quantum,farhi2014quantum_applied}.

\begin{figure}[t]	
	\centering
	\includegraphics[width=\columnwidth]{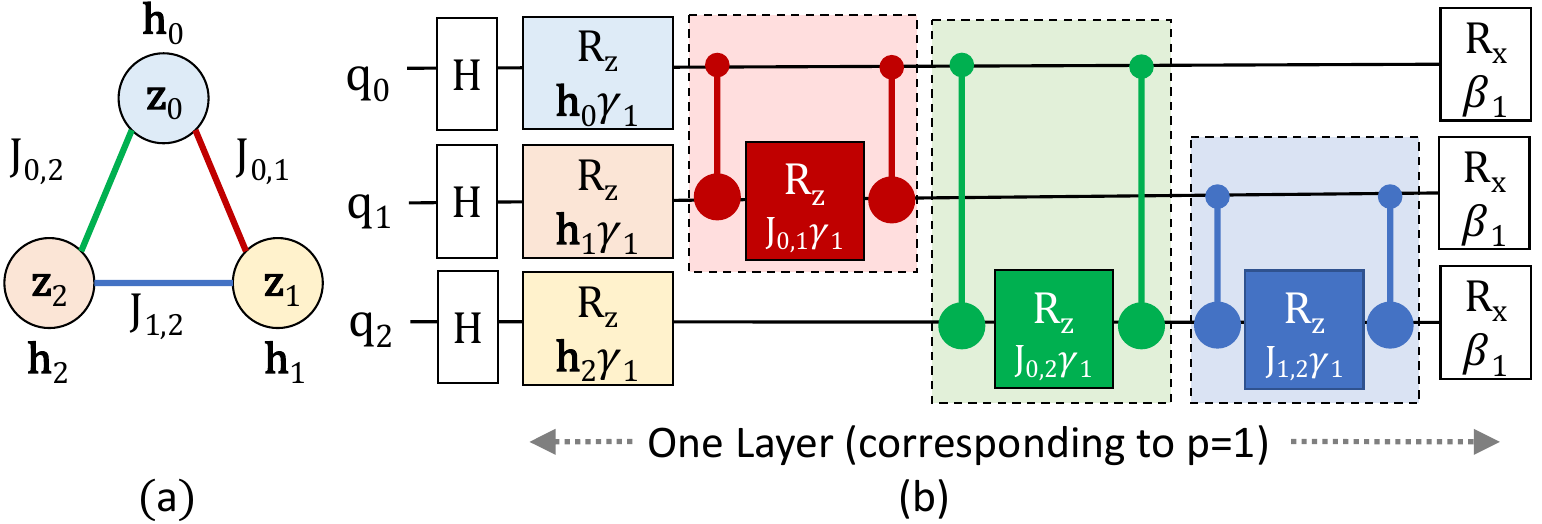}
    \caption{		
		 A QAOA example:
(a)	graph representation of a three-qubit Hamiltonian; 
	(b) QAOA circuit with one layer. 
	}	
	\label{fig:qaoa_example}
\end{figure}

For example, to represent the Max-Cut problem using Ising Hamiltonian, we add the $\mathbf{z}_i\mathbf{z}_j$ term to the Ising Hamiltonian for every edge between nodes $i$ and $j$ and 
use $J_{ij}$ to indicate the weight of the edge between node $i$ and $j$, as shown in Figure \ref{fig:qaoa_example}(a). 
If running the QAOA algorithm and finding optimum value yield $\mathbf{z}_i\mathbf{z}_j = -1$, it means that nodes $i$ and $j$ are in two separate cuts in the QAOA solution for that Max-Cut problem. 
Alternately, $\mathbf{z}_i\mathbf{z}_j = 1$ implies the nodes belong to the same cut. 
QAOA is computationally universal~\cite{morales2020universality,lloyd2018quantum}and can be used for tackling a wide range of real-world applications. 
However, formulating practical problems as minimizing Equation~\eqref{eqn:Ising} in general is nontrivial~\cite{ayanzadeh2021equal,ayanzadeh2020multi,ayanzadeh2022quantum,ayanzadeh2020ensemble,ayanzadeh2020reinforcement,ayanzadeh2019quantum,lucas2014ising}.

Solving Equation~\eqref{eqn:Ising} on a quantum computer requires a QAOA circuit with 
(1)~$N$ qubits, 
(2)~one single-qubit gate for each $\mathbf{z}_{i},$ and
(3)~two two-qubit gates, in addition to one single-qubit gate, for each $\mathbf{z}_{i}\mathbf{z}_{j}$ term. 
This comprises a single layer in the circuit. Thus, the number of single and two-qubit operations in a QAOA circuit scales with the number of nodes and edges in the problem graph, respectively. 
For example, Figure~\ref{fig:qaoa_example}(b) shows the QAOA circuit for an Ising Hamiltonian that represents the graph in Figure~\ref{fig:qaoa_example}(a). 
QAOA circuits can consist of multiple layers ($p$). 
In a QAOA circuit, $\mathbf{z}_{i}$ indicates the result of measuring qubit $i$. In quantum computing, when one measures qubits $\ket{0}$ and $\ket{1}$ on $\mathbf{z}$ basis, the outcomes are the eigenvalues of operator $\mathbf{z}$, which are $+1$ and $-1$, respectively. 

QAOA circuits are parametric circuits, as shown in Figure~\ref{fig:qaoa_example}(b), and involves searching for the optimal values of the parameters $\gamma_1$ and $\beta_1$ (which corresponds to the angles of the single-qubit rotation gates) using a classical optimizer. 
QAOA applications are run in a variational mode where the output distributions of the circuits are used by the classical optimizer to adjust the next round of parameters and this training process continues until the optimizer converges. The solution of the problem is inferred from the output distribution of the QAOA circuit using the optimal parameters. 
Demonstrating quantum advantage---outperforming the state-of-the-art classical optimization techniques---at a practical scale 
requires running QAOA circuits with (at least) some hundreds of qubits~\cite{guerreschi2019qaoa} and several layers~\cite{basso2021quantum}.

\subsection{Impact of Hardware Errors on QAOA Circuits}
Near-term quantum devices are noisy and suffer from high error rates. 
For example, qubits do not retain their information beyond a few micro-seconds on most existing systems~\cite{sycamoredatasheet}, a phenomenon referred to as \textit{decoherence}. 
Also, imperfections in quantum gate and measurement operations cause programs to encounter computational errors when programs are executed on NISQ devices. 
In particular, CNOT operations are the dominant sources of errors on most existing NISQ systems with an average error-rate of about 1\%~\cite{sycamoredatasheet}. They also incur long latencies (about 400 ns on existing IBMQ systems) and therefore, increases the probability of decoherence. This limits the fidelity of QAOA circuits as the problem size grows due to an increase in the number of two-qubit operations. 

\begin{figure}[h]	
	\centering
	\includegraphics[width=0.95\columnwidth]{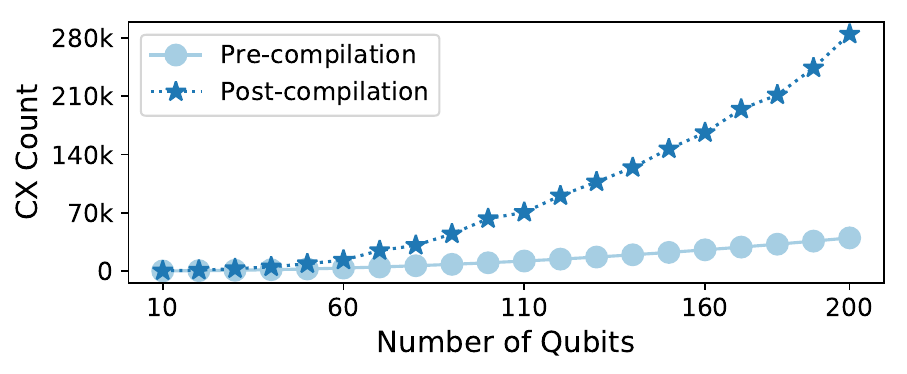}
	\vspace{-0.1in}
    \caption{		
	Impact of SWAPs on fully-connected QAOA graphs on a grid qubit architecture.
	}	
   	\vspace{-0.1in}
	\label{fig:cx_count}
\end{figure}

NISQ devices also suffer from limited connectivity where qubits are only connected to some of their nearest neighbors. To execute CNOT operations between unconnected qubits, NISQ compilers introduce SWAP operations. 
A SWAP is a sequence of three CNOT operations that interchanges the state of two qubits. 
Thus, they enable compilers to overcome the limited connectivity of NISQ devices by routing non-adjacent qubits next to each other. 
Unfortunately, SWAPs increase the total number of CNOT operations and depth of the circuits, making them even more vulnerable to errors. 
For example,  Figure~\ref{fig:cx_count} shows that the number of CNOTs of a compiled program increases up-to 14X even for small programs. 
The problem compounds when QAOA circuits with multiple layers must be executed to accurately estimate the solution of a problem~\cite{basso2021quantum} as the number of CNOT operations and depth increase, exacerbating the impact of hardware errors~\cite{dei2006exact,hadlock1975finding,harrigan2021quantum}.

\begin{figure*}[t!]	
	\centering
	\includegraphics[width=\linewidth]{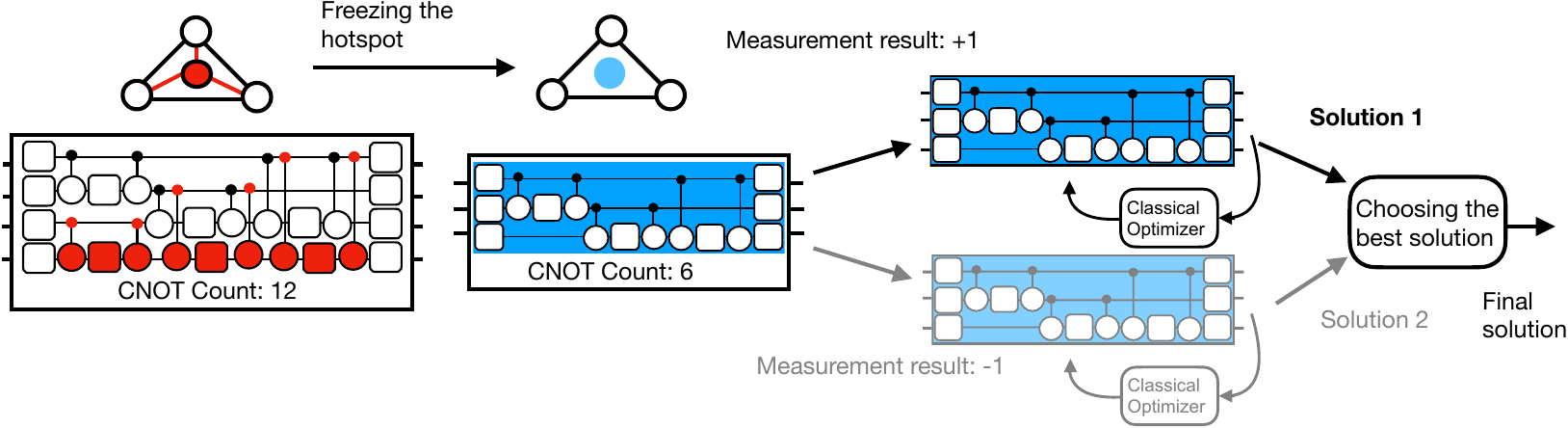}
    \caption{Overview of the FrozenQubits framework when one qubit is frozen. After freezing the hotspot qubit, we have two sub-problems with smaller quantum circuits. Due to the symmetricity of the search space of most QAOA problems, we can skip half of the sub-problems and find the best solution for only the remaining half sub-problems; see section \ref{sec:symmetricity}.}
	\label{fig:FrozenQubits_1q_overview}
\end{figure*}

\section{FrozenQubits: Freeze the HotSpots}
We present {\em FrozenQubits}, an application-level software framework, to increase the fidelity of QAOA applications by \textit{freezing} or skipping hotspot nodes and solving sub-problems with smaller more reliable quantum circuits. In this section, we provide our insight and explain the design of FrozenQubits.

\subsection{Insight and Goal: Leverage Application Characteristics to Improve Fidelity of QAOA Circuits}
The goal of this work is to decrease the impact of two-qubit gate errors, 
which are the main source of errors in QAOA circuits run on near-term quantum computers. 
To achieve this goal, we design a software framework that reduces the number of two-qubit gates of the QAOA circuit by leveraging application-level properties of real-world problem graphs.

\begin{tcolorbox}[colback=green!12]
\textbf{Insight}: In real-world problem graphs, not all nodes contribute equally to the QAOA circuit.  
\end{tcolorbox}

\vspace{0.05in}
\noindent \textbf{Power-Law Graphs:}
In most natural and artificial graphs, the degree of connectivity---number of edges connected to a node---of a small number of nodes are much higher than the rest of nodes. 
In other words, the degree distributions of most real-world application graphs follow \textbf{\emph{Power-Law}} distribution~\cite{agler2016microbial,barabasi1999emergence,clauset2016colorado,gamermann2019comprehensive,goh2002classification,house2015testing,jeong2001lethality,mislove2007measurement,pastor2015epidemic}. 
For example, if we consider the graph for friendship relations on a social network, we observe that the majority of people in many friendship interactions have only a few connections to other people, whereas some people, such as influencers, are \textit{hotspots} and are connected to many others in the network~\cite{sfGraph}. 
We make similar observations in many other real-world problem graphs. 
For example, Table~\ref{table:power-law} shows various power-law graphs in real-world problem domains along with the application of QAOA to solve them on near-term quantum computers.

\begin{table}[h]
\caption{Power-Law Graphs for Real-World Problems and The Usage of QAOA to Solve Them.}\vspace{0.15in}
    \centering
    \setlength{\tabcolsep}{0.12cm} 
    \renewcommand{\arraystretch}{1.3}
    \begin{tabular}{|l|l|l|l|}
        \hline
        \multirow{2}{*}{Domain} & Sub-Domain & Power-Law & QAOA  \\
        & & Example & Application\\
        \hline
        \hline
        \multirow{2}{*}{Transportation} & Vehicle Routing & \cite{akbarzadeh2018look,berliant2018scale,mori2020common} & \cite{9774961,bentley2022quantum,https://doi.org/10.48550/arxiv.2110.06799}\\
        \cline{2-4}
        & Supply Chain & \cite{hou2018does,suo2018exploring} & \cite{bmw_supplychain,bentley2022quantum}\\
        \hline
        \multirow{2}{*}{Biology} & Protein Folding & \cite{magner2015origin,qian2001protein,sawada2006structural} & \cite{emani2021quantum,https://doi.org/10.48550/arxiv.1810.13411,robert2021resource}\\
        \cline{2-4}
        & DNA Sequences &  \cite{buldyrev2006power,costa2019analysis,peng1995statistical} & \cite{bulancea2019quantum,sarkar2021quaser} \\
        \hline
        Finance and & Portfolio & \cite{ait2015robust,elmerraji2021optimal,wu2021fractal} & \cite{baker2022wasserstein,Barkoutsos_2020,https://doi.org/10.48550/arxiv.2207.10555,egger2020quantum}\\
        Economics & Optimization & &\\
        \cline{2-4}
          & Auctions & \cite{kahng2004emergence} & \cite{egger2020quantum}\\
        \hline
    \end{tabular}\vspace{0.15in}
    \label{table:power-law}
\end{table}

In QAOA, the number of edges for a node determines the number of two-qubit CNOT gates, contributed by that node to the quantum circuit. 
Therefore, the hotspots contribute to a significantly larger number of CNOT gates in the QAOA circuits,
compared to the other nodes, for real-world applications.

\subsection{Overview of FrozenQubits}
The number of two-qubit gates in QAOA circuits scales with the number of edges. In real-world problem graphs, some nodes are hotspots and contribute more to the number of edges.
Using these characteristics of graphs, we propose \emph{FrozenQubits}, an application-level software framework that improves the  fidelity of QAOA applications.
FrozenQubits freezes the hotspot qubits meaning it removes those qubits and their associated two-qubit gates from the circuit. This divides the problem into smaller sub-problems of which the associated quantum circuits are less vulnerable to error. Figure \ref{fig:FrozenQubits_1q_overview} shows an overview of FrozenQubits scheme for freezing one qubit.

\subsection{Freezing Qubits and Defining Sub-Problems}

When one \emph{freezes} a qubit, it means that the qubit, along with the gates connected to it, are eliminated from the circuit. 
By freezing the qubit we assume that we measure that qubit in $\mathbf{z}$-basis. 
There are two possibilities for the result of measurements in $\mathbf{z}$-basis, $+1$ or $-1$. 
For each of these possibilities, we need to define a sub-Hamiltonian. 
Let qubit $k$ be the one frozen by the FrozenQubits. 
The two sub-Ising Hamiltonians $H^{\mathbf{z}_k = +1}_Z$ in Equation \eqref{eqn:Hamiltonianz+1} and $H^{\mathbf{z}_k = -1}_Z$ in Equation \eqref{eqn:Hamiltonianz-1} are obtained by substituting $\mathbf{z}_k$ in the original Ising Hamiltonian $H_Z$ with $+1$ and $-1$, respectively. 
Table \ref{table:notations} summarizes the notations used in defining the original Hamiltonian and sub-Hamiltonians.

\begin{table}[b]
\caption{Notations of Ising Hamiltonians for a Given Problem and Sub-Problems After Freezing}
    \centering
    \setlength{\tabcolsep}{0.12cm} 
    \renewcommand{\arraystretch}{1.3}
    \begin{tabular}{|c|l|}
        \hline
        Notation & Definition  \\
        \hline
        \hline
        \multirow{1}{*}{$H_Z$} & $\sum_{i=0}^{N-1}{\mathbf{h}_i \mathbf{z}_i} +
		\sum_{i=0}^{N-1}{ \sum_{j=i+1}^{N-1}{J_{ij} \mathbf{z}_i \mathbf{z}_j} }
		+\text{\textit{offset}}$, \\
        & shows the Ising Hamiltonian of the original problem\\
        \hline
        $\mathbf{z}_i$ & Result of measuring qubit $i$ in $\mathbf{z}$-basis; $\mathbf{z}_i \in \{-1,+1\}$\\
        \hline
        \multirow{1}{*}{$J_{ij}$} & Coefficient of quadratic term $\mathbf{z}_i\mathbf{z}_j$ in $H_z$. In graph\\
        & representation, it indicates the weight of the edge\\
        & between node $i$ and node $j$. \\
        \hline
        \multirow{1}{*}{$\mathbf{h}_i$} & Coefficient of linear term $\mathbf{z}_i$ in $H_z$. In graph\\
        & representation, it indicates the weight of the node $i$. \\
        & In Max-Cut problem, the weight of all nodes are zero,\\
        & $\mathbf{h}_i = 0, \forall i$. \\
        \hline
        $H^{\mathbf{z}_k = +1}_Z$ & Ising Hamiltonian of a sub-problem for measurement \\
        & result $+1$ after freezing qubit $k$\\
        \hline
        
        $H^{\mathbf{z}_k = -1}_Z$ & Ising Hamiltonian of a sub-problem for measurement \\
        & result $-1$ after freezing qubit $k$\\
        \hline
        
        \multirow{1}{*}{$\mathbf{h}^{\mathbf{z}_k = +1}_i$} & Linear coefficient in $H^{\mathbf{z}_k = +1}_Z$, after freezing qubit $k$,\\
        & for node $i$ which is equal to  $\mathbf{h}_i	+  J_{k,i} +   J_{i,k}$.\\
        \hline
        
        \multirow{1}{*}{$\mathbf{h}^{\mathbf{z}_k = -1}_i$} & Linear coefficient in $H^{\mathbf{z}_k = -1}_Z$, after freezing qubit $k$,\\
        & for node $i$ which is equal to  $\mathbf{h}_i	-  J_{k,i} -   J_{i,k}$.\\
        \hline
        
        \multirow{1}{*}{$\text{{offset}}^{\mathbf{z}_k = +1}$} & Offset of $H^{\mathbf{z}_k = +1}_Z$, after freezing qubit $k$, which is\\
        & equal to $\text{\textit{offset}}+ \mathbf{h}_k$\\
        \hline
        
        \multirow{1}{*}{$\text{{offset}}^{\mathbf{z}_k = -1}$} & Offset of $H^{\mathbf{z}_k = -1}_Z$, after freezing qubit $k$, which is\\
        & equal to $\text{\textit{offset}}-\mathbf{h}_k$\\
        \hline
    \end{tabular}
    \label{table:notations}
\end{table}

\begin{equation}\label{eqn:Hamiltonianz+1}
	H^{\mathbf{z}_k = +1}_Z =  \sum_{\substack{i=0 \\ i \neq k}}^{N-1}{\mathbf{h}^{\mathbf{z}_k = +1}_i \mathbf{z}_i} +\sum_{\substack{i=0 \\ i \neq k}}^{N-1}{ \sum_{\substack{j=i+1 \\ j \neq k}}^{N-1}{J_{ij} \mathbf{z}_i \mathbf{z}_j} }
		+\text{{offset}}^{\mathbf{z}_k = +1}
\end{equation}

\begin{equation}\label{eqn:Hamiltonianz-1}
	H^{\mathbf{z}_k = -1}_Z =  \sum_{\substack{i=0 \\i \neq k}}^{N-1}{\mathbf{h}^{\mathbf{z}_k = -1}_i  \mathbf{z}_i}  +
		\sum_{\substack{i=0 \\ i \neq k}}^{N-1}{ \sum_{\substack{j=i+1 \\ j \neq k}}^{N-1}{J_{ij} \mathbf{z}_i \mathbf{z}_j} }
		+\text{{offset}}^{\mathbf{z}_k = -1}
\end{equation}

As shown in Equations \eqref{eqn:Hamiltonianz+1} and \eqref{eqn:Hamiltonianz-1}, each sub-problem corresponds to exactly one half of the state-space of $H_Z$---one sub-problem for substituting $\mathbf{z}_k$ with +1 and another sub-problem for substituting $\mathbf{z}_k$ with -1.
Repeating the substituting process for remaining qubits on the resulting sub-problems will partition the state-space of $H_Z$ into much smaller sub-spaces. 
More specifically, freezing $m$ qubits will partition the state-space of $H_Z$ into $2^m$ sub-spaces, and any of the resulting sub-problems will have $N - m$ variables and accordingly their associated sub-circuit will have $N - m$ qubits.

\begin{figure}[ht]
	\centering
	\includegraphics[width=0.8\linewidth]{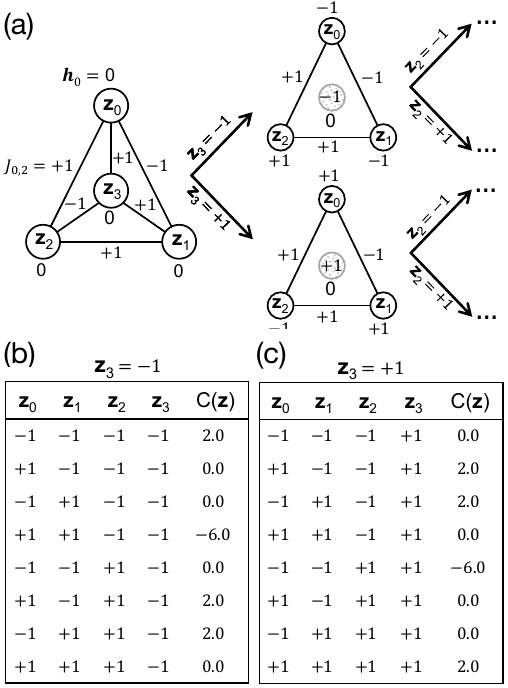}	
    \caption{
	Example of freezing a qubit for an Ising Hamiltonian with four qubits.
	Substituting $\mathbf{z}_3$ with +1 and -1 results in two sub-problems with three spin variables in each. 
	}
	\label{fig:fixing_qubits}
\end{figure}

From a graph representation viewpoint, substituting $\mathbf{z}_k$ drops all edges that are connected to $\mathbf{z}_k.$ 
Figure \ref{fig:fixing_qubits}(a) illustrates the process of freezing qubits using the graph representation of an example problem with four qubits. 
Substituting $\mathbf{z}_3$ with +1 and -1 results in two sub-problems with three qubits. 
Figure \ref{fig:fixing_qubits}(b) shows the state-space of sub-problems when we substitute $\mathbf{z}_3$ with -1 and +1. 
The union of all sub-spaces is identical to the state-space of the original problem of interest.

For each sub-Hamiltonian, we need to run the classical optimization step on each associated QAOA circuit. All the sub-Hamiltonians of the original problem have the same quadratic forms. However, they vary in terms of offsets and linear coefficients. Therefore, the general structure of the QAOA circuit for all sub-Hamiltonians is similar and they only differ in terms of angles of the rotation gates. These angles are trainable parameters learned during the optimization of QAOA.

Note that unlike the given problem where all linear coefficients were zero, the resulting sub-problems have non-zero linear coefficients (i.e., $\mathbf{h}_k \neq 0)$.
As shown in Figure \ref{fig:qaoa_example}, every linear term of $H_Z$ (with a linear coefficient of $\mathbf{h}_k)$ corresponds to an $R_z$ gate in each layer of the QAOA circuit. 
However, $R_z$ gates are software gates and do not impact the fidelity.

\subsection{Optimizing Number of Qubits to Freeze: A\\Fidelity-Cost Trade-off}
The performance of FrozenQubits depends on the number of qubits frozen. 
However, there exists a trade-off between the fidelity improvement and the quantum cost of freezing qubits. 
While freezing an increased number of qubits allow us to drop a larger number of CNOT operations and design smaller sub-circuits that execute with greater fidelity, 
the quantum cost of executing the sub-circuits grow exponentially. 
More specifically, the quantum cost of freezing ${m}$ qubits is $O\left( {2^m} \right)$.  

Finding the optimum number of qubits to freeze is non-trivial. 
To overcome this challenge, FrozenQubits leverages the insight that for most real-world applications that follow Power-law distribution, the number of dropped edges per qubit decreases quickly for the hotspots (due to higher connectivity) but the pace of CNOT reduction decreases for additional nodes beyond the hotspots. 
Thus, freezing only a limited number of nodes is sufficient. 
We confirm our insights using experiments on real systems and observe that freezing additional nodes beyond a certain point has diminishing returns (we defer the discussion to Section~\ref{sec:cost_fidelity_tradeoff}). 
Thus, FrozenQubits can leverage circuit properties such as CNOT counts and depth to determine the number of qubits to freeze for a given application. 
As FrozenQubits is a scalable framework, we leave it up to the user to select the number of qubits to freeze. Our default design considers dropping up to two qubits.

\subsection{Which Qubits to Freeze?}

For a given QAOA problem, FrozenQubits can choose $m$ qubits to freeze from $^NC_m$ possibilities. 
However, instead of randomly selecting qubits from all possibilities, FrozenQubits selects the $m$ qubits corresponding to the hotspots in the problem graph. 
The insight is that freezing hotspots in real-world problem graphs allows FrozenQubits to drop the maximum number of CNOT operations in the QAOA circuit. 
Moreover, hotspots also contribute to a significantly larger number of SWAP operations compared to other nodes. 
Therefore, freezing hotspots allow FrozenQubits to also reduce the SWAP overheads to a much larger extent compared to other nodes.

\subsection{Decoding Outcomes}
After finding the optimum value of each sub-problem, we need to find the final solution of the original problem of interest. 
FrozenQubits partitions the state-space of the input problem into smaller sub-spaces and explores them independently via running multiple smaller QAOA programs. 
Every sub-problem corresponds to one of the sub-spaces while each includes $N - m$ qubits. 
Therefore, we can find the solutions with the best objective value, the lowest cost value, by just calculating the minimum of the solutions of the sub-problems. 
FrozenQubits, in contrast to previous works, has no postprocessing and no cost for finding the final solution after solving the sub-problems, except finding the minimum over the solutions of the sub-problems.

\subsection{Tackling the Overheads of FrozenQubits} 
When we freeze $m$ qubits, we will have $2^m$ smaller quantum circuits to train. 
This growth in the number of circuits increases the overheads because 
1)~we need to compile these circuits for execution on a quantum computer and 
2)~we need to run all circuits independently and infer their outputs to find the solution for the primary problem of interest. 
Here, we discuss how we tackle these overheads using the characteristics of sub-problem Hamiltonians.

\subsubsection{Reducing the Compilation Overhead:} \label{sec:compilation_reduction}
Freezing ${m}$ qubits results in $2^m$ separate QAOA sub-problems. 
To find the solution of each sub-problems, we run optimization steps on their associated QAOA circuits. 
However, these circuits only vary in terms of angles of rotation gates. 
This is because all the Hamiltonians of sub-problems have the same terms, and they only differ in terms of different coefficients and offset values. 
Therefore, we only compile one \emph{template circuit}, and edit the resulting compiled circuit for generating executable circuits for all sub-problems. 
Editing the compiled circuit means embedding $\mathbf{h}_i$ and $J_{ij}$ into the angles of the corresponding $R_z$ rotations. 
This approach significantly reduces the compilation overhead of FrozenQubits.

\subsubsection{Pruning Sub-Problems:} \label{sec:symmetricity}
When we freeze a qubit and create two sub-problems by substituting the frozen qubit $\mathbf{z}$ value in the Hamiltonian with $-1$ and $+1$,  the state-space of these two sub-problems can be symmetric. 
For example in Figure~\ref{fig:fixing_qubits}(b) and Figure~\ref{fig:fixing_qubits}(c), values of $C(z)$ is symmetric with respect to $\mathbf{z}$. This means flipping all $\mathbf{z}$ values of any row from Figure \ref{fig:fixing_qubits}(b)---i.e., $+1 \to -1$ and $-1 \to +1$---corresponds to a row in Figure \ref{fig:fixing_qubits}(c), and vice versa. 
Here we demonstrate that this symmetricity appears in all Hamiltonians with all zero linear coefficients. 

When all linear coefficients of an Ising Hamiltonian, shown in Eq. \eqref{eqn:Ising}, is set to zero $(\mathbf{h}_i=0$ for $i=0, 1, \dots, n-1)$ 
we have 
\[	
	C({\mathbf{z}}) =  \sum_{i=0}^{n-1}{ \sum_{j=i+1}^{n-1}{J_{ij} \mathbf{z}_i \mathbf{z}_j} }. 
\]
Note that we have omitted the ``offset`` term since (as a constant) it does not have any impact on the shape or structure of the problem landscape. 
Since $\mathbf{z}_i \in \{ {-1,+1} \},$ 
$C({\mathbf{z}})$ can be re-written as 
\[	
	C({\mathbf{z}}) =  \sum_{i=0}^{n-1}{ \sum_{j=i+1}^{n-1}{\pm J_{ij} }} 
\]
where the sign of $J_{ij}$ only depends on values of $\mathbf{z}_i$ and $\mathbf{z}_j.$
In other words, when both $\mathbf{z}_i$ and $\mathbf{z}_j$ have the same value, their product will be +1. Otherwise, their product will be -1. 
While flipping all variables will change values of $\mathbf{z}_i$ and $\mathbf{z}_j$ individually, their product value will remain unchanged. Thus, 
\[
	C(\mathbf{z}) = C(-{\mathbf{z}}).
\]
If $\mathbf{z}^*$ is a global minimum of an Ising model  with zero linear coefficients, 
$\mathbf{-z}^*$ is also a global minimum of the same Ising Hamiltonian. 
Moreover, we can conclude that the number of global minimums in an Ising model with zero linear coefficients is even.

FrozenQubits leverages this symmetricity to mitigate the quantum cost. 
When a qubit is frozen and all linear coefficients of the parent problem are zero, FrozenQubits executes QAOA steps on just one of the sub-problems. 
After training circuit parameters for this sub-problem, one flips the $\mathbf{z}$ values of the solution to construct the output distribution of the other sub-problem. 
This pruning significantly mitigates the quantum cost of FrozenQubits.

\subsection{Scalability of FrozenQubits}
Let ${m}$ be the number of qubits to freeze, $N$ be the number of qubits, ${s}$ be the number of distinct outcomes in an output distribution, and $|J|$ be the number of quadratic terms in $H_z.$

\vspace{0.05in}
\noindent \textbf{Quantum complexity} 
The quantum resource utilization in FrozenQubits scales exponentially with the number of skipping qubits, $\mathbf{O\left( {2^m} \right)}$.
However, we can eliminate a significant number of sub-problems and substantially subside the quantum overhead of FrozenQubits, without compromising the performance of the primary QAOA application (we discussed it in Section \ref{sec:symmetricity}). 
Note that ${m}$ does not scale with $N$ and for power-law graphs $m\ll N$.

\vspace{0.05in}
\noindent \textbf{Circuit compilation complexity:}
Assuming that all sub-problems are run on the same quantum computer, FrozenQubits only compiles one \emph{template circuit}; 
accordingly, the compilation complexity of FrozenQubits is $\mathbf{O\left( {1} \right)}$. 
This takes significantly less time compared to compiling the QAOA circuit for the baseline.

\vspace{0.05in}
\noindent \textbf{Time complexity:} 
The complexity of required tasks to be performed on a classical computer depends on the complexity of different components. 
The complexity for identifying the top ${m}$ hotspot nodes is $O\left(N + m\log{m} \right),$ assuming that the adjacency list of the graph representing $H_Z$ is available. 
The complexity order of forming the adjacency list in the worst case (i.e., fully connected graphs) is $O(N^2).$ 
A node is connected to at most $N-1$ other nodes; therefore, freezing ${m}$ nodes scales with $O(mN),$ and forming sub-problems scales with $O\left( {ms2^m} \right).$
Decoding every outcome (whith $N-m$ bits) to the state-space of the original problem Hamiltonian is $O(m).$
Hence, inferring the final solution is $O \left( {s2^m (m + N + |J|) } \right).$ 
For problems at a practical scale, $m \ll N \ll s.$ Therefore, classical time complexity scales with the order of $\mathbf{O\left( {sN^2} \right)}$, excluding the circuit compilation time that is reduced significantly with increasing ${m}$.

\vspace{0.05in}
\noindent \textbf{Memory complexity:}
To identify and freeze hotspot qubits, FrozenQubits use the adjacency list representation of the input problem graph which has the space complexity of $O\left( {N^2} \right).$ 
Since sub-problems are independent, decoding the output distribution has the space complexity of $O\left( {sN} \right).$ 
For QAOA applications at a practical scale, we expect that $N \ll s;$ thus, the overall space complexity of FrozenQubits is $\mathbf{O\left( {sN} \right)}$.

\subsection{Comparison with Prior Works That Use Sub-Circuits}
CutQC is a prior studies that divides a quantum circuit into smaller sub-circuits~\cite{bravyi2016trading,peng2020simulating,tang2021cutqc}. 
While CutQC is applicable to any quantum circuit, it works best when there are limited connections between qubits in the input quantum circuit. However, this is not true for QAOA and other variational quantum algorithms. 
Moreover, CutQC is bottle-necked by exponentially complex post-processing that scales with the number of qubits. On the other hand, FrozenQubits freezes only some of the hotspots (up to two in our default design) to create a limited number of circuits and does not incur such post-processing costs. Table~\ref{tab:comparisonwithcutqc} compares FrozenQubits with CutQC.
\begin{table}[h]
\caption{Comparison of FrozenQubits and CutQC.}
    \centering
    \setlength{\tabcolsep}{0.12cm} 
    \renewcommand{\arraystretch}{1.3}
    \begin{tabular}{|l|l|l|l|l|}
        \hline
        \multirow{2}{*}{Design} & \multirow{2}{*}{Application} & \multicolumn{3}{c|}{Overheads}   \\
        \cline{3-5}
        & & Compile & Quantum  & Post-process  \\
        \hline\hline
        \multirow{2}{*}{CutQC} & \multirow{2}{*}{Generic} & \multirow{2}{*}{Linear} & \multirow{2}{*}{Linear}& Exponential \\
        & & & & (in qubits) \\
        \hline
        \multirow{2}{*}{FrozenQubits} & \multirow{2}{*}{QAOA }& \multirow{2}{*}{$O(1)$}& Exponential$^*$ & Polynomial\\
        & & & (in $m)$ & \\
        \hline
    \end{tabular}
       \begin{tablenotes}
   \footnotesize
   \item $^*$\textit{Our default FrozenQubits design only freezes up-to $m=2$ qubits. For real-world applications, freezing a few hotspots is sufficient for FrozenQubits to be effective. Please see Section~\ref{sec:cost_fidelity_tradeoff} for an analysis on this. }
    \end{tablenotes}    
    \label{tab:comparisonwithcutqc}
\end{table}
\newpage
\section{Methodology}

\subsection{Benchmarks}

\begin{figure}[ht]		
	\centering
	\vspace{-2pt}
	\includegraphics[width=\columnwidth]{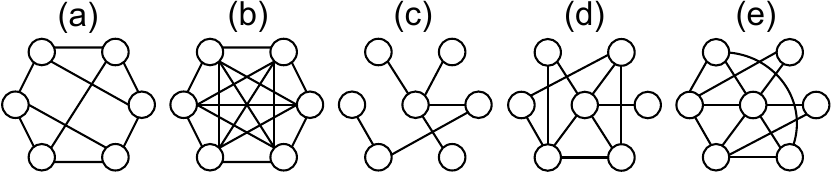}
    \caption{		
		 Random graphs: (a) 3-regular, (b) SK model, (c) BA $(d_{\text{BA}}=1)$, (d) BA $(d_{\text{BA}}=2)$, and (e) BA $(d_{\text{BA}}=3)$.
	}	
	\label{fig:benchmark_graphs}
\end{figure}

We study FrozenQubits on three types of graphs: (1)~Power-law, (2)~3-regular, and (3)~fully-connected or SK-model graphs. 
While most real-world problems follow Power-law distribution, existing quantum computers cannot run problems at such scale as they involve hundreds of qubits. 
So instead, we generate smaller Power-law graphs that mimic the characteristics of real-world problems but can be run on real systems available today. 
To generate Power-law graphs, we use the widely accepted \textit{Barabasi—Albert (BA)} algorithm~\cite{albert2005scale,barabasi1999emergence,barabasi2000scale,gray2018super,kim2022sparsity,lusseau2003emergent,wang2019complex,zadorozhnyi2012structural,zbinden2020embedding}. 
BA graphs are associated with a preferential attachment factor $d_{\text{BA}}$ that controls the the density of the graphs. 
We generate Power-law graphs using the BA algorithm for $d_{\text{BA}}=1$ as prior studies show that $d_{\text{BA}}=1$ can capture the dynamics of most real-world systems~\cite{clauset2016colorado}. 
To study FrozenQubits on denser graphs, we use BA graphs corresponding to $d_{\text{BA}}=2$ and 3. 
For all the graphs, the edge weights are randomly drawn from $\{-1,+1\},$ and all node coefficients $(\mathbf{h}_i)$ are set to zero~\cite{das2008colloquium,harrigan2021quantum}. 
Figure \ref{fig:benchmark_graphs} shows five samples of the benchmark graphs used in this study.

\subsection{Baseline and Experimental Platform}
\textbf{Baseline}: We run the QAOA circuits with the optimal circuit parameters (determined from simulations) on the NISQ hardware for 100K trials. This approach is consistent with prior works on compilers for QAOA~\cite{alam2020circuit}. Note that prior works show that additional trials do not improve application fidelity once the distribution saturates after a certain point~\cite{jigsaw}. To compile each circuit, we use IBM's Qiskit tool-chain with noise-adaptive routing and the highest optimization level 3. 

\vspace{0.05in}
\noindent \textbf{Quantum hardware.}: We use eight IBMQ systems with 27—127 qubits: Washington, Brooklyn, Montreal, Auckland, Toronto, Mumbai, Hanoi, and Cairo.

\subsection{Figure of Merit} 

We evaluate the application fidelity of QAOA circuits using
Approximation Ratio Gap (ARG) from prior works~\cite{alam2020circuit,das2021jigsaw,herrman2022multi}, as defined in Equation~\eqref{eqn:ARG}. 
\vspace{-0.1in}
\begin{equation}
	\text{ARG} = 100 \times \left|{ \frac{EV_{ideal} - EV_{real}}{EV_{ideal}}}\right|
	\label{eqn:ARG}
\end{equation}
where $EV_{ideal}$ and $EV_{real}$ denote the expected values of the QAOA circuit on an ideal simulator and the real quantum machine, respectfully. 
$\text{ARG} \in [0, +\infty]),$ and lower is better. 

\newpage
\section{Evaluation}

\subsection{FrozenQubits on Power-Law graphs}
We evaluate FrozenQubits using Barabasi Albert (BA) graphs~\cite{barabasi1999emergence} that represent most power-law graphs~\cite{clauset2016colorado}. 

\subsubsection{Impact of Number of CNOTs and Circuit Depth}

\begin{figure}[h]		
	\vspace{-0.1in}
	\captionsetup[subfigure]{position=top} 
	\centering	
	\subfloat[]{
		\includegraphics[width=0.5\columnwidth]{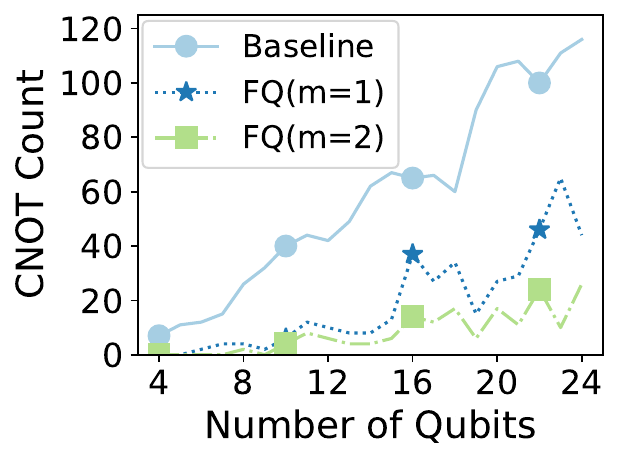}
	}\hspace*{-0.8em}		
	\subfloat[]{
		\includegraphics[width=0.5\columnwidth]{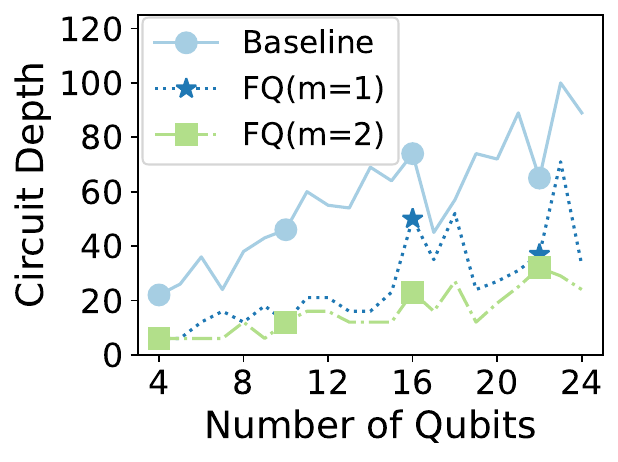}
	}
	\caption{
		(a) CNOT counts and (b) Depth 
		of QAOA and FrozenQubits (FQ) circuits, 
		when $m=1$ and 2 qubits are frozen. }
	\label{fig:FQ(1)_circ_properties}
\end{figure}

Figure~\ref{fig:FQ(1)_circ_properties}(a) shows that FrozenQubits reduces the number of CNOTs by 3.13x on average when only one qubit is frozen. 
The CNOT count reduces by 7.19x on average when two qubits are frozen. 
Figure~\ref{fig:FQ(1)_circ_properties}(b) shows the impact of FrozenQubits on circuit depth. 
Freezing a single qubit reduces circuit depth by 2.23x on average. 
Note that the CNOT count and circuit depth include the overheads from SWAP operations. 
Freezing two qubits reduces depth by 3.65x on average. 

\subsubsection{Impact of FrozenQubits on Application Fidelity} \hfill
\label{sec:app_fidelity}

\begin{figure}[h]
	\centering
	\vspace{-0.1in}
	\includegraphics[width=\columnwidth]{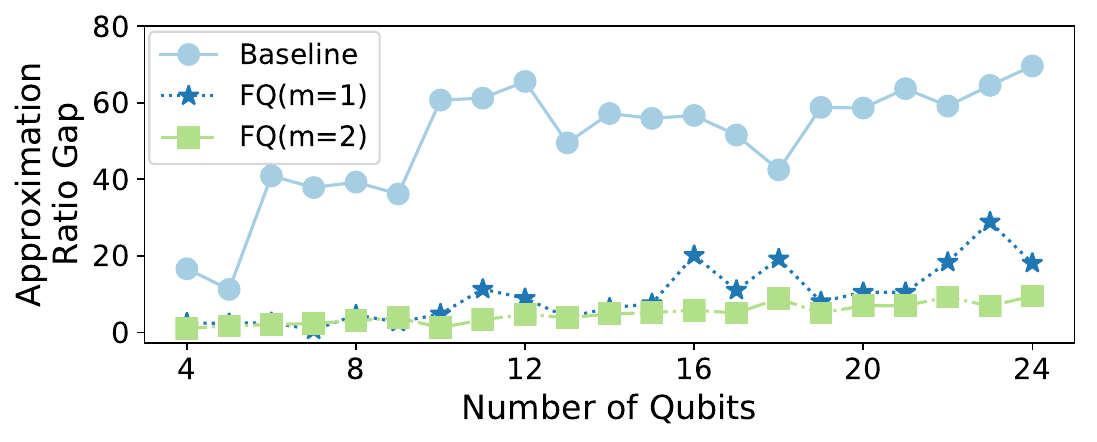}	
    \caption{
	Approximation Ratio Gap (ARG) on IBM-Montreal 
	using standard QAOA and FrozenQubits (FQ) for $m=1$ and $2$. 
	}	    
	\label{fig:expr_BA_1_ARG}
\end{figure}

\noindent Figure~\ref{fig:expr_BA_1_ARG} shows that FrozenQubits improves the Approximation Ratio Gap (ARG) of the power-law (BA) QAOA circuits by 6.75x on average and by up-to 47.04x when $m=1$ qubit is frozen. 
Freezing 2 qubits improves the ARG by 11.29x on average and by up-to 57.14x.

From Figure~\ref{fig:expr_BA_1_ARG}, we also observe that the fidelity of the baseline reduces (increasing ARG) rapidly with the circuit size. 
On the contrary, the ARG reduces at a much slower rate with increasing problem size for FrozenQubits. 

As FrozenQubits exploits the symmetry of the search space, it does not incur any quantum cost when only a single node is frozen, 
and freezing two qubits requires twice the quantum resources. 
For more information regarding how leveraging the symmetry of search space reduces the overheads of FrozenQubits, see Section \ref{sec:symmetricity}.

\subsubsection{Cost and Fidelity Trade-off} 
\begin{figure}[h]	 
	\vspace{-0.1in}
	\captionsetup[subfigure]{position=top} 
	\centering	
	\subfloat[]{
		\includegraphics[width=0.48\columnwidth]{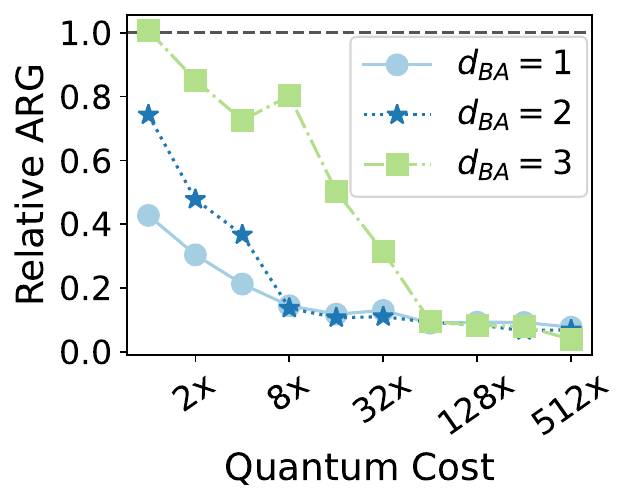}
	}\hspace*{-0.8em}
	\subfloat[]{
		\includegraphics[width=0.52\columnwidth]{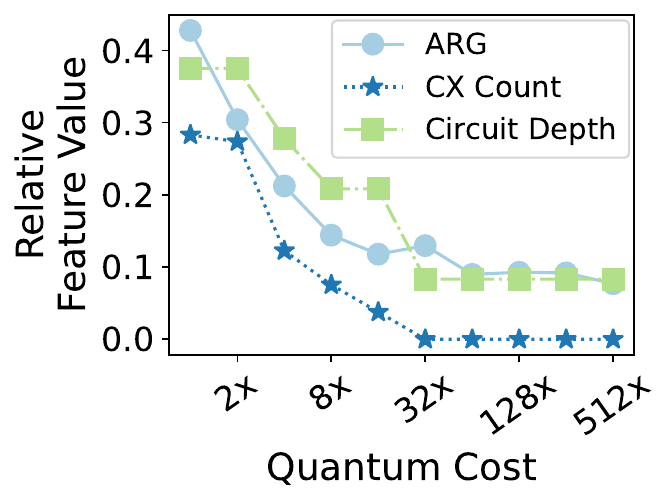}
	}%
    \caption{	 
		Trade-off between quantum cost and fidelity.}
	\label{fig:FQ_10}
\end{figure}

\hfill \label{sec:cost_fidelity_tradeoff}
\noindent There exists a trade-off between the fidelity improvement from freezing additional qubits and the quantum cost of running FrozenQubits. 
While freezing more qubits improves the fidelity of QAOA circuits, it simultaneously increases the quantum cost of FrozenQubits exponentially. 
However, our evaluations show that freezing additional qubits has diminishing returns after a certain point. 
For example, Figure~\ref{fig:FQ_10}(a) shows that the improvement in ARG saturates after freezing seven qubits.

Therefore, to make a trade-off, we must:
(1)~determine the quantum budget; 
and (2)~roughly estimate the point (i.e., number of qubits to freeze) where the trend in improving the fidelity plateaus. 
The quantum budget is user-specific and depends on various factors such as the availability of quantum resources and specific requirements of the underlying applications. 
To estimate the number of qubits to be frozen, we can use circuit properties such as number of CNOTs and depth. 
For example, Figure~\ref{fig:FQ_10}(b) shows that these circuit parameters can accurately capture the application fidelity trends of the QAOA circuits.

\subsubsection{FrozenQubits on Dense Power-law (BA) Graphs} 
\begin{figure}[h]		
	\captionsetup[subfigure]{position=top} 
	\centering	
	\vspace{-0.1in}
	\subfloat[$d_{\text{BA}}=2$]{
		\includegraphics[width=0.5\columnwidth]{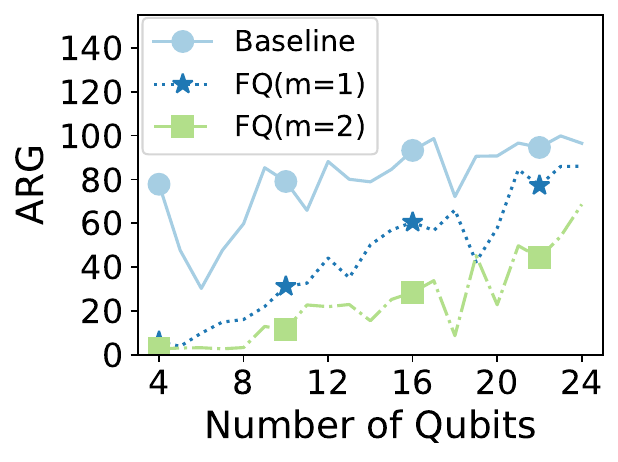}
	}\hspace*{-0.8em}		
	\subfloat[$d_{\text{BA}}=3$]{
		\includegraphics[width=0.5\columnwidth]{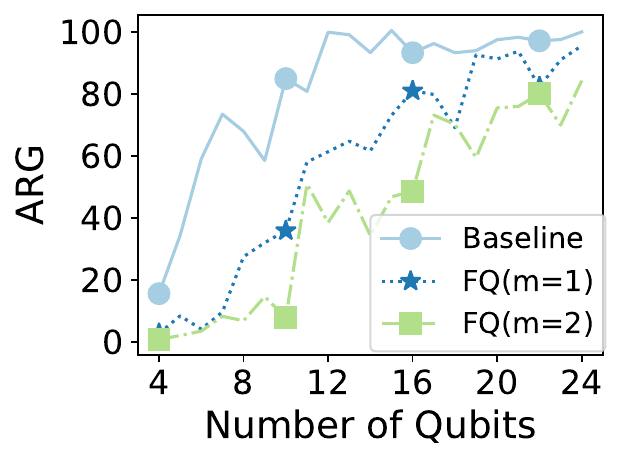}
	}
	\caption{
		Approximation Ratio Gap (ARG) of denser 
		BA graphs
		on IBM Montreal. 
	}
	\label{fig:denser_BA}
\end{figure}

\hfill
\label{sec:densebaresults}
\noindent To study the impact of FrozenQubits on denser power-law graphs, we use BA graphs with $d_{\text{BA}}=2$and 3. A higher $d_{\text{BA}}$ corresponds to a denser graph.  
Figure~\ref{fig:denser_BA} shows that for dense graphs ($d_{\text{BA}}=2$), when one qubit is frozen, FrozenQubits improves ARG by 1.76x on average and by up-to 12.8x. 
Even for very dense graphs, corresponding to $d_{\text{BA}}=3$, FrozenQubits improves the ARG by 1.43x on average and by up-to 14.1x.
Freezing two qubits enhances the performance even further, as shown in Figure~\ref{fig:denser_BA}. 

\subsection{FrozenQubits on Regular Graphs}
\begin{figure}[h]
	\captionsetup[subfigure]{position=top} 
	\centering	
	\subfloat[]{
		\includegraphics[width=0.49\columnwidth]{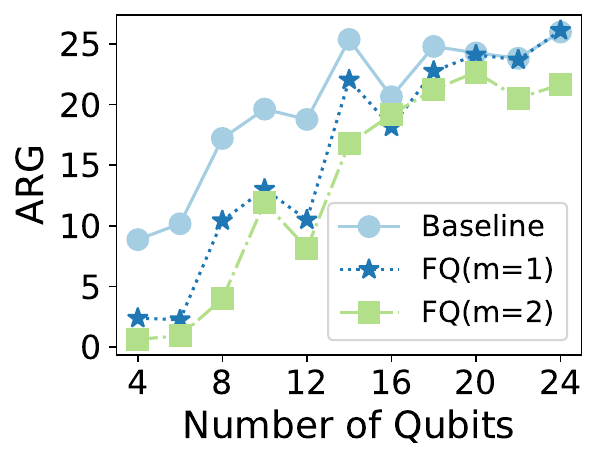}
	}\hspace*{-0.8em}		
	\subfloat[]{
		\includegraphics[width=0.51\columnwidth]{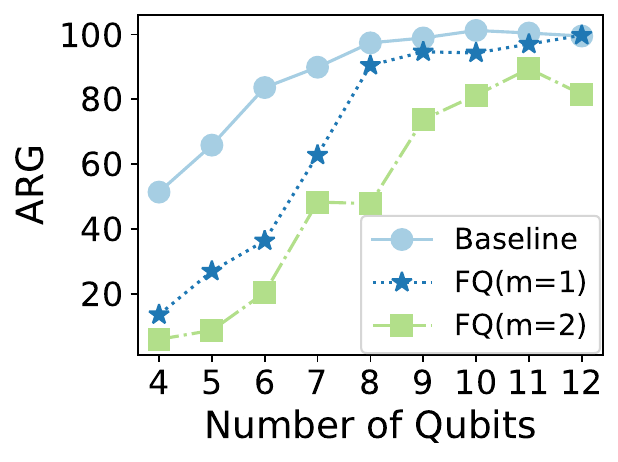}
	}
	\caption{
		Approximation Ratio Gap (ARG) 
		of (a) 3-regular graphs and (b) SK model 
		on IBM-Montreal.
	}	    
	\label{fig:regular_graphs} 
\end{figure}

We study FrozenQubits on 
(1) 3-regular graphs and 
(2) fully connected graphs (or SK model)~\cite{sherrington1975solvable,harrigan2021quantum,basso2021quantum}. 
Figure~\ref{fig:regular_graphs} shows that FrozenQubits improves the ARG of QAOA for 3-regular graphs by 1.25x on average (and by up to 4.52x). For SK-model graphs, the ARG improves by 1.28x on average and by up to 3.79x when a single qubit is frozen. 
Freezing two qubits improves the effectiveness of FrozenQubits further.

\begin{figure*}[b]
	\centering	
	\includegraphics[width=0.7\textwidth]{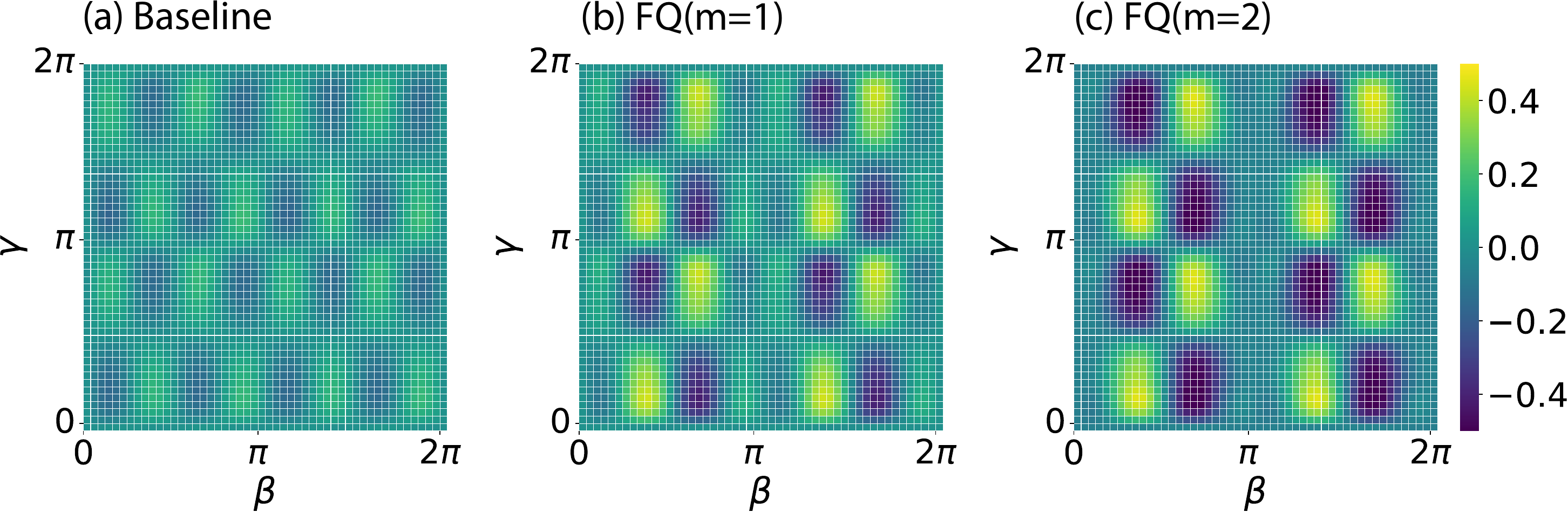}	
	\caption{
		Classical optimizer landscape of AR in Equation~\eqref{eq:AR} for 
		(a)~baseline, (b) FrozenQubits with $m=1,$ and (c) FrozenQubits with $m=2,$
		on IBMQ-Auckland for a 20-qubit power-law graph. 
		Note that the landscapes are not identical as the search space of the Hamiltonians in $\text{FQ}(m=1)$ and $\text{FQ}(m=2)$ correspond to half and quarter of the search space, respectively. 
	}	
    
	\label{fig:hitmap}	
\end{figure*}

\subsection{Impact of Resources on Training QAOA} 

\textbf{Executing the Baseline for More Trials:} 
The fidelity of the baseline does not improve once the output distribution saturates after few thousands of trials due to correlated errors
~\cite{jigsaw}. 
Recent QAOA studies from Google uses 25K trials for programs using up-to 25 qubits, whereas our baseline is executed for 100K trials. 

\vspace{0.05in}
\noindent \textbf{Executing the Baseline for More Iterations:}
The performance of QAOA depends on (1)~the fidelity of the circuits and (2)~the ability of the optimizer to tune the circuit parameters using the output of the quantum circuits. 
Unfortunately, noisy outputs impact the training as the circuits lose their sensitivity to parameters changes~\cite{harrigan2021quantum,sung2020using}. 

To understand the impact of FrozenQubits on the training process of QAOA, we run the baseline and FrozenQubits for a 20-qubit power-law graph across a $50\times50$ optimization landscape
and compute the approximation ratio (AR) as:
\begin{equation}
\label{eq:AR}
	AR = \frac{\text{Expected Value}}{C_{\text{min}}}
\end{equation}
where $C_{\text{min}}$ is the global minima. 
Each point denotes a unique combination of two circuit parameters ($\gamma$ and $\beta$). 
AR $\in [-\infty ,1]$ and is maximum when all outcomes in the output distribution are global optima~\cite{harrigan2021quantum,sung2020using}. 
Note that we perform this analysis on the grid to compare the baseline and FrozenQubits on the entire landscape as opposed to a specific path traversed by the optimizer.

Figure~\ref{fig:hitmap} shows that the landscape of the baseline is much more blurred due to noise, compared to FrozenQubits. 
Thus, even if the baseline is executed for more iterations, it may still not improve the quality of the solution as the sensitivity of the quantum circuits is lowered by noise. 
However, the reduced noise in FrozenQubits enables it to sharpen the gradients in the parameter landscape considerably. 
Therefore, circuits are more sensitive to changes in parameters in FrozenQubits, aiding the training of QAOA.

\subsection{Sensitivity of FrozenQubits to Machines}
\begin{figure}[h]
	\centering	
	\includegraphics[width=\columnwidth]{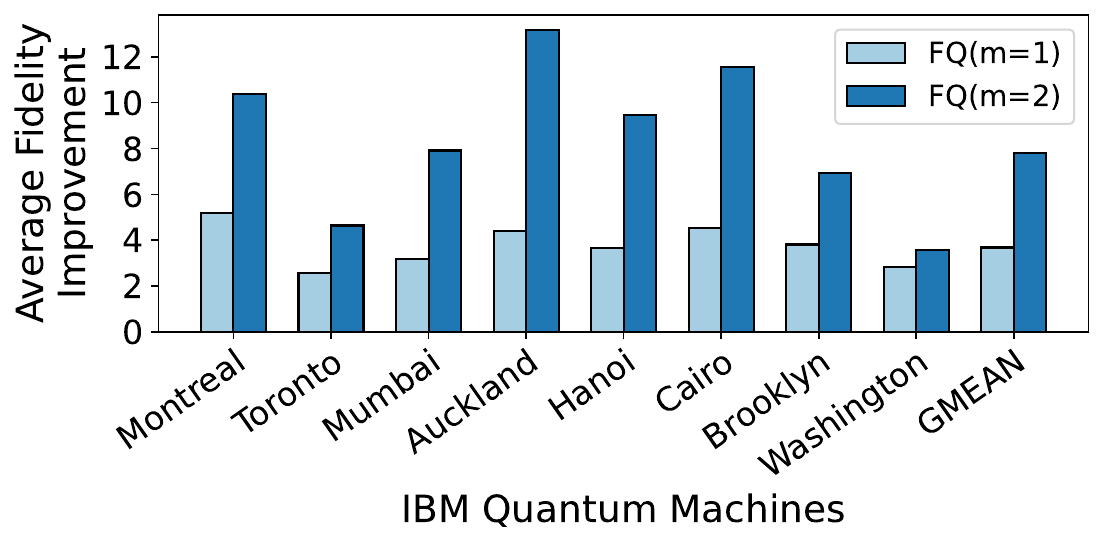}
    \caption{
	Average ARG improvement across IBMQ-systems.
	}
	\label{fig:expr_BA_1_relative_ARG_all_machines}
\end{figure}

Figure~\ref{fig:expr_BA_1_relative_ARG_all_machines} shows that freezing one qubit across different machines improves the mean ARG of QAOA by 3.69x on average and by up to 5.20x. The ARG improves by 7.8x on average and by up to 13.16x when two qubits are frozen.

\newpage
\section{FrozenQubits at Practical Scale} \label{sec:FQ_at_scale}
Demonstration of quantum advantage using QAOA requires solving problems at the scale of several hundreds of qubits~\cite{guerreschi2019qaoa}. Unfortunately, we cannot run such applications on existing quantum computers. Even a state-of-the-art device such as 53-qubit Google Sycamore cannot run QAOA benchmarks beyond 23 qubits~\cite{harrigan2021quantum}. To evaluate the impact of FrozenQubits at practical scale, we study its performance using 500-qubit random power-law QAOA circuits on a $50\times50$ grid. 

\subsection{Impact on Number of CNOTs} 
\begin{figure}[h]
	\centering	
	\includegraphics[width=\columnwidth]{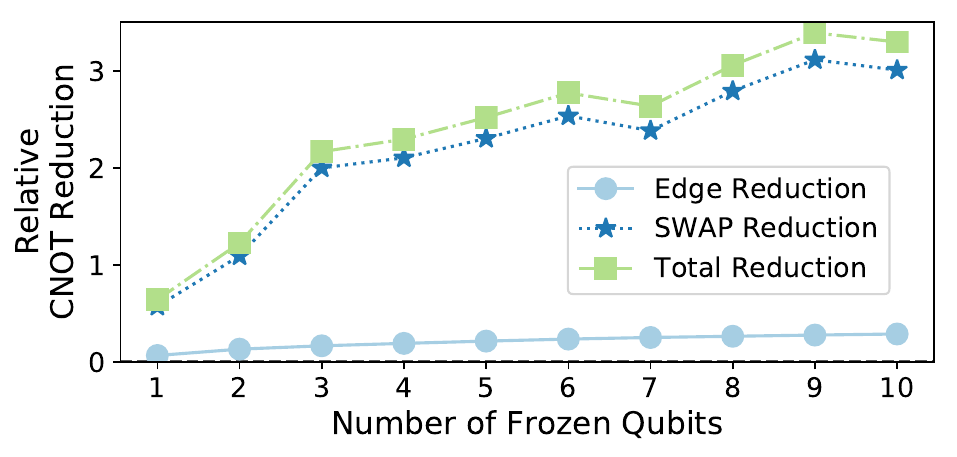}
    \caption{
	Relative CNOT reduction on BA benchmarks with $d_{\text{BA}} = 1.$ 
	Higher is better.
	}    
	\label{fig:at_scale_rel_CNOT_BA_1}
\end{figure}

Figure~\ref{fig:at_scale_rel_CNOT_BA_1} shows the reduction in CNOT counts post-compilation for a 500-qubit circuit. 
Increasing the number of frozen qubits ($m$) reduces the number of CNOTs. For example, freezing ten qubits results in 65.94\% CNOT reduction, at the cost of training 512 circuits independently. Further analysis shows that  91.47\% of this reduction corresponds to reduced number of SWAP operations. Moreover, as FrozenQubits skip hotspot nodes, SWAP reduction on average provides 10.19x higher contribution to the total CNOT reduction, compared to reduction of CNOTs in the QAOA circuit due to the dropped edges corresponding to the skipped nodes (at the circuit before compilation). 
We make similar observations even on denser BA graphs, as shown in Figure \ref{fig:at_scale_rel_CNOT_BA_all}(a). 

\subsection{Impact on Circuit Depth}
\begin{figure}[h]
	\captionsetup[subfigure]{position=top} 
	\centering	
	\subfloat[]{
		\includegraphics[width=0.5\columnwidth]{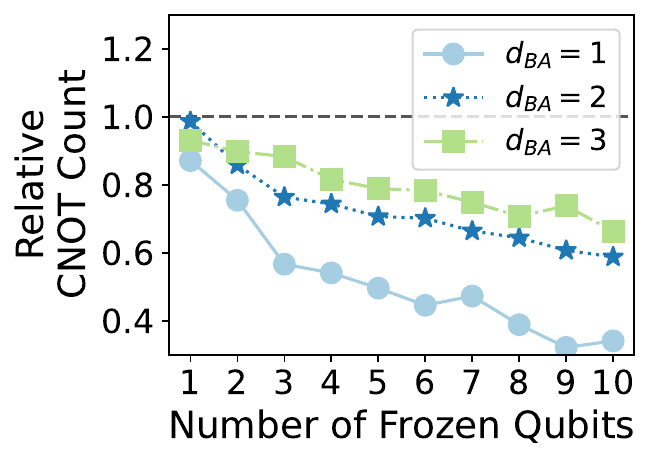}
	}\hspace*{-0.8em}		
	\subfloat[]{
		\includegraphics[width=0.5\columnwidth]{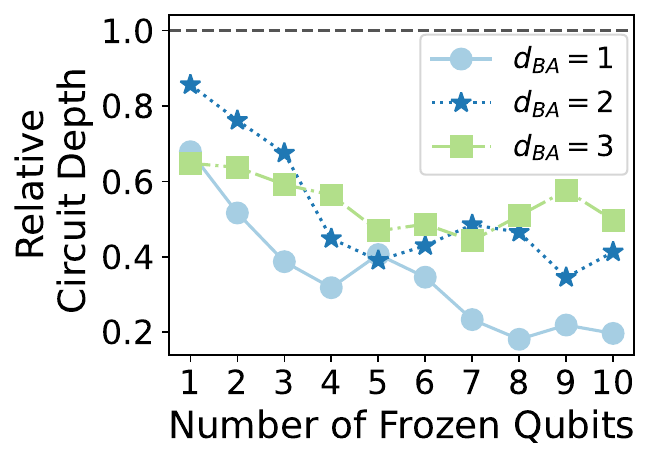}
	}
	\caption{		
	Relative (a) CNOT counts and (b) circuit depth,
		for BA benchmarks with $d_{\text{BA}} = 1, 2$ and 3. Lower is better. 
}	
	\label{fig:at_scale_rel_CNOT_BA_all}
\end{figure}

Figure~\ref{fig:at_scale_rel_CNOT_BA_all}(b) 
shows that FrozenQubits reduces circuit depth of BA benchmarks by 1.47x--5.25x on average when the number of qubits frozen increase from one to ten.

\subsection{Impact on Expected Probability of Success} 
\begin{figure}[h]
	\centering	
	\includegraphics[width=\columnwidth]{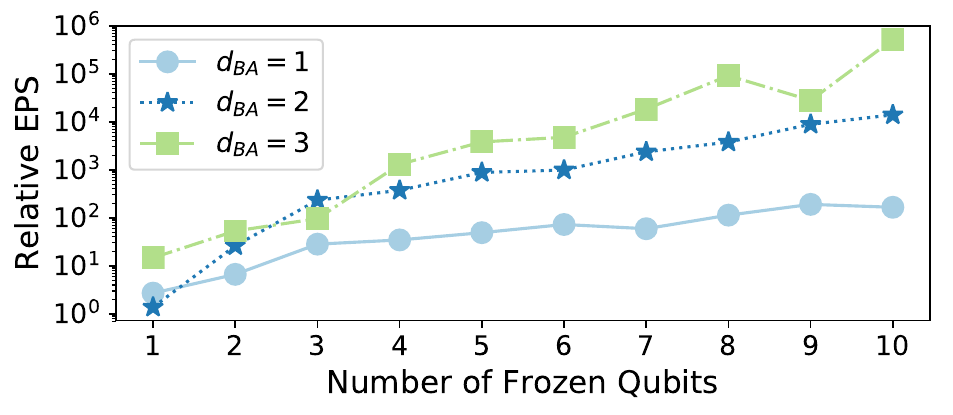}
    \caption{
	Expected Probability of Success (EPS) comparison.
	}
	\label{fig:eps_figure}
\end{figure}

Running such large problems on existing quantum computers to evaluate the performance of FrozenQubits is infeasible. 
Instead, we compute the \emph{Expected Probability of Success (EPS)} using an optimistic error model, 
where we assume 0.1\% CNOT error-rate and 
0.5\% readout error-rate. 
We also assume a 500$\mu$ seconds decoherence time.
EPS is the probability that gate and measurement operations remain error-free and qubits remain free from decoherence. 
It is widely used in evaluating the performance of NISQ compilers on large programs~\cite{alam2020circuit,jigsaw,nishio,tannu2022hammer}. 
Figure~\ref{fig:eps_figure} shows that FrozenQubits improves the EPS by 404x on average and by up to 515,900x. 
 
\subsection{Impact on Compilation Time} \label{}

\begin{figure}[h]	
	\captionsetup[subfigure]{position=top} 
	\centering	
	\subfloat[]{
		\includegraphics[width=0.53\columnwidth]{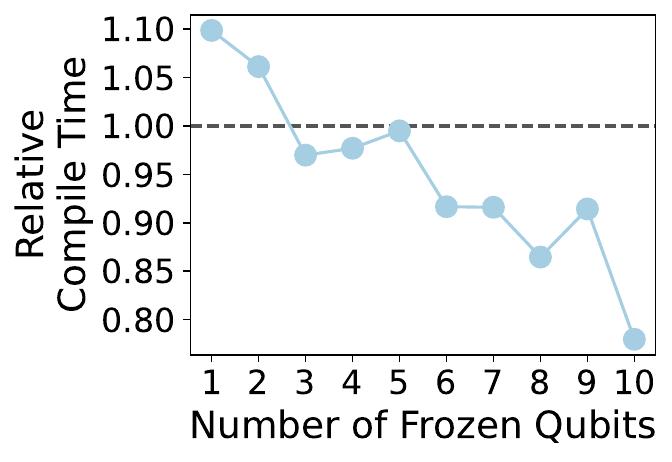}
	}\hspace*{-0.8em}		
	\subfloat[]{
		\includegraphics[width=0.47\columnwidth]{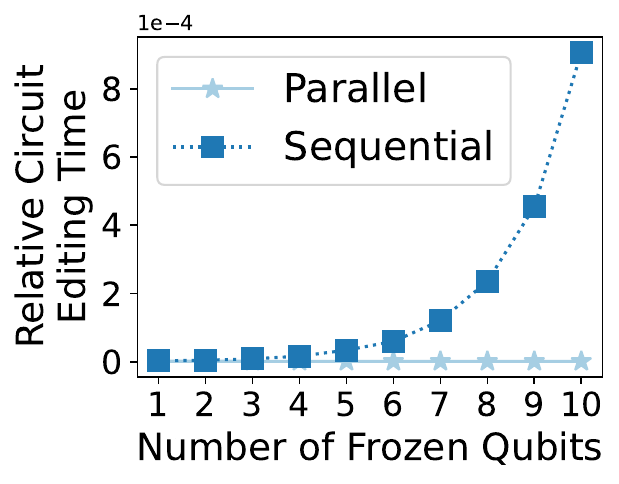}
	}
	\caption{		
		(a) Relative compilation time. (b) Time required to generate all executables for FrozenQubits, relative to the time required to compile the baseline.
		}	
	\label{fig:compilationtime}
\end{figure}

As the sub-circuits from FrozenQubits have fewer operations compared to the baseline and require fewer SWAPs, their compilation time is lower. But, FrozenQubits requires $2^m$ sub-circuits when $m$ qubits are frozen and each of them must be compiled into an executable. 
However, our approach reduces the compilation overhead by compiling only one of the sub-circuits and generating the remaining executables by editing it (sequentially or in parallel). 
This approach reduces that compilation overheads of FrozenQubits. For example, Figure~\ref{fig:compilationtime}(a) shows that freezing ten qubits decreases the compilation time by 22.06\%. This reduction is significantly higher than the overall time required to generate all the $O\left( {2^m} \right)$ executables, as shown in Figure~\ref{fig:compilationtime}(b). 

\subsection{End-to-End Workflow Runtime Analysis} \label{sec_latency}
Quantum computers are accessed via cloud services~\cite{ibmqcloud,AmazonBraKet,castelvecchi2017ibm,IBMQ,MicrosoftAzure} 
and the overall runtime of a quantum circuit depends on several factors such as (1)~queuing delays, (2)~execution mode, and (3) execution time on the NISQ device. 
The cloud management software frameworks differ between providers and are evolving to provide flexibility to the users~\cite{das2019case,ravi2021quantum,ravi2021adaptive}. 
For example, users can access devices in \textit{dedicated} or \textit{shared} mode~\cite{AmazonBraKet}.  
Similarly, some device providers allow launching multiple circuits simultaneously as part of a single cloud job, for instance, up to 900 circuits at once on IBMQ systems~\cite{IBMQ}. 
To capture the diverse execution models and fairly compare the runtime of the baseline and FrozenQubits, we use an analytical model described by Equation~\eqref{eq:analyticalmodel}, 
where $I$ is the number of QAOA iterations, $\tau$ is the number of trials, $t_{NISQ}$ is the execution time for each trial, $N_{batch}$ is the number of batches required, $\Delta_{cloud}$ is the cloud access latency, $\Delta_{opt}$ is the latency of the classical optimizer, $\delta_{compile}$ is the compilation latency, and $\delta_{pp}$ is post-processing time needed (if any). 


\begin{align}
	T = \delta_{compile} &+ I\times N_{batch} \times \left({ \tau \times t_{NISQ}+\Delta_{cloud}}\right) \nonumber \\
		&+ \delta_{opt} + \delta_{pp}
	\label{eq:analyticalmodel}
\end{align}

\begin{figure}[h]
	\centering	
	\includegraphics[width=\columnwidth]{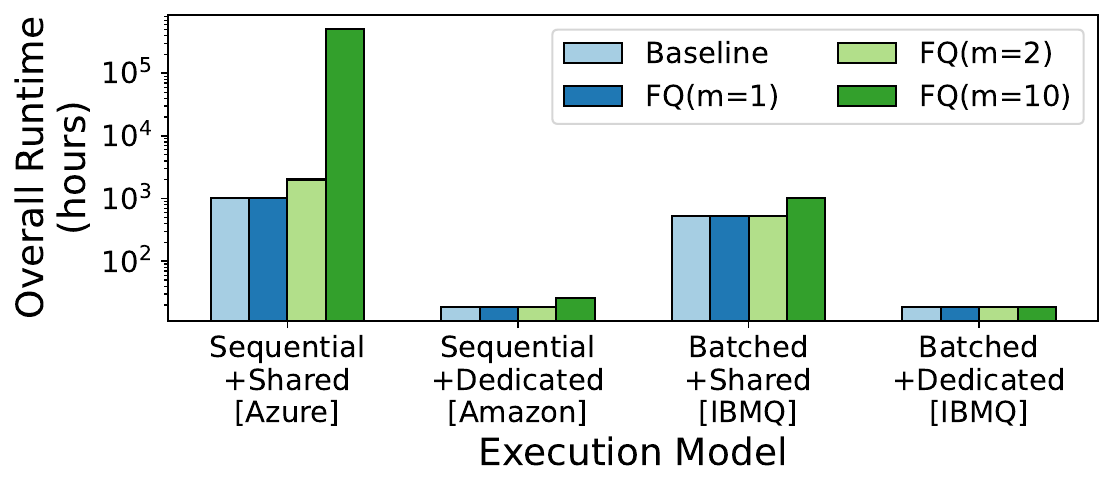}	
    \caption{
	Overall Runtime comparison.
	}
    \vspace{-0.15in}
	\label{fig:latency}
\end{figure}

For our analysis, we assume (1)~two execution modes: no-batching (such as Rigetti devices~\cite{AmazonBraKet}) 
and batching up-to 900 circuits as on IBMQ systems~\cite{IBMQ}; 
and (2)~two device access modes: shared with $\Delta_{cloud}=30$ minutes  and 
dedicated with $\Delta_{cloud}=0$. 
By default, we assume both the baseline and FrozenQubits run $\tau=25K$ trials~\cite{harrigan2021quantum} 
per circuit and it requires $t_{NISQ}=$ 1 millisecond to run a trial. 
We also assume $\Delta_{opt}=1$ minute is the optimizer latency for an iteration, 
and by default we need $I=1000$ iterations per circuit. 
Both the baseline and FrozenQubits compile a single circuit only once and we assume a latency of $\delta_{compile}=2$ hours. 
Lastly, we assume post-processing time of FrozenQubits, $\delta_{pp}=1$ minute. 
Figure~\ref{fig:latency} compares the end-to-end runtime of the baseline, default FrozenQubits that freezes up to two qubits, and FrozenQubits freezing 10 qubits. Note that, freezing only some of the hotspots is sufficient for FrozenQubits to be effective.

The end-to-end runtime for both the baseline and FrozenQubits depends on the execution model. 
The batching of circuits allow FrozenQubits to launch all sub-circuits in an iteration simultaneously, reducing the runtime. 
We use a simplified model to estimate the runtime and it will be lower if a user improves the throughput by multi-programming NISQ devices~\cite{9407180,das2019case} or running circuits on multiple devices~\cite{stein2022eqc,ravi2021quantum}. 

\newpage
\section{Related Work} \label{sec:related_work}
Software policies to improve the fidelity of NISQ applications is an active area of research. 
We can classify these techniques as generic and QAOA-specific policies. 
Generic policies do not use the domain knowledge of the underlying application, and try to improve the fidelity of NISQ computers by: 
(1)~better compiling quantum circuits~\cite{
barron2020measurement,
das2021adapt,
das2021jigsaw,
gokhale2020optimized,
kwon2020hybrid,
li2018tackling,
noiseadaptive,
murali2020software,
patel2022geyser,
shi2019optimized,
smith2021error,
tannu2019not,
xie2022suppressing}; and 
(2)~postprocessing output distributions~\cite{bravyi2020mitigating,matrixmeasurementmitigation,patel2020veritas,tannu2022hammer}.
These policies are orthogonal to our proposed technique, and one may combine them with FrozenQubits. 
Approximating quantum circuits~\cite{patel2022quest} and circuit cutting techniques~\cite{tang2021cutqc} can improve the fidelity of NISQ computers. 
However, applying these techniques to QAOA circuits with hotspots is nontrivial.

QAOA-specific policies try to leverage the domain knowledge of the underlying problem for improving the fidelity of QAOA applications~\cite{lao20222qan,li2022paulihedral}. 
Reordering Pauli terms of Hamiltonians can result in QAOA circuits with lower depth and fewer CNOTs~\cite{alam2020circuit}. 
However, increasing the number of CNOTs that are run in parallel can escalate the crosstalk~\cite{xie2022suppressing}. 
Partial compilation of circuits in variational algorithms at the pulse level~\cite{gokhale2019partial} can shorten the execution time of QAOA but has the overheads of dealing with custom pulses.

\section{Conclusion}	
We propose \emph{FrozenQubits}, an application-level software framework for boosting the fidelity of Quantum Approximate Optimization Algorithm (QAOA).
It leverages the insight that most natural and artificial graphs follow Power-law distribution. 
FrozenQubits freezes hotspot qubits and intelligently partitions the state-space of the problem into several smaller sub-spaces such that the corresponding QAOA sub-circuits are significantly less robust to hardware errors on NISQ devices.
To subside the quantum complexity of FrozenQubits, we define and prove a new theorem that eliminates running a considerable number of QAOA processes for sub-problems without losing the guarantee of the recovery of the exact solution. 
Our evaluations using 5,300 QAOA circuits on eight IBMQ computers show that FrozenQubits improves the fidelity of QAOA circuits by up to 57.14x compared to the baseline.

\section*{Acknowledgements}
We thank Swamit Tannu and Nicolas Delfosse for helpful discussions and comments. 
We also thank Suhas Vittal for his inputs on the error modelling and editorial comments.
Ramin Ayanzadeh was supported by the NSF Computing Innovation Fellows (CI-Fellows) program. 
This work was funded in part by EPiQC, an NSF Expedition in Computing, under grant CCF-1730449.  Poulami Das was funded by the Microsoft Research PhD Fellowship. 
This research used resources of the Oak Ridge Leadership Computing Facility at the Oak Ridge National Laboratory, which is supported by the Office of Science of the U.S. Department of Energy under Contract No. DE-AC05-00OR22725.
This research was supported in part through research cyberinfrastructure resources and services provided by the Partnership for an Advanced Computing Environment (PACE) at the Georgia Institute of Technology, Atlanta, Georgia, USA

\bibliographystyle{ACM-Reference-Format}
\bibliography{references}


\begin{thebibliography}{116}


\ifx \showCODEN    \undefined \def \showCODEN     #1{\unskip}     \fi
\ifx \showDOI      \undefined \def \showDOI       #1{#1}\fi
\ifx \showISBNx    \undefined \def \showISBNx     #1{\unskip}     \fi
\ifx \showISBNxiii \undefined \def \showISBNxiii  #1{\unskip}     \fi
\ifx \showISSN     \undefined \def \showISSN      #1{\unskip}     \fi
\ifx \showLCCN     \undefined \def \showLCCN      #1{\unskip}     \fi
\ifx \shownote     \undefined \def \shownote      #1{#1}          \fi
\ifx \showarticletitle \undefined \def \showarticletitle #1{#1}   \fi
\ifx \showURL      \undefined \def \showURL       {\relax}        \fi
\providecommand\bibfield[2]{#2}
\providecommand\bibinfo[2]{#2}
\providecommand\natexlab[1]{#1}
\providecommand\showeprint[2][]{arXiv:#2}

\bibitem[bmw({[n.\,d.]})]%
        {bmw_supplychain}
 \bibinfo{year}{[n.\,d.]}\natexlab{}.
\newblock \bibinfo{title}{BMW's supply chain makes a quantum leap}.
\newblock
  \bibinfo{howpublished}{\url{https://www.cips.org/supply-management/analysis/2021/may/bmw-supply-chain-makes-a-quantum-leap/}}.
\newblock
\newblock
\shownote{Accessed: 2022-07-27}.


\bibitem[sfG({[n.\,d.]})]%
        {sfGraph}
 \bibinfo{year}{[n.\,d.]}\natexlab{}.
\newblock \bibinfo{title}{How Network Math Can Help You Make Friends}.
\newblock
  \bibinfo{howpublished}{\url{https://www.quantamagazine.org/how-network-math-can-help-you-make-friends-20180820/}}.
\newblock
\newblock
\shownote{Accessed: 2022-07-27}.


\bibitem[ibm(2021)]%
        {ibmqcloud}
 \bibinfo{year}{2021}\natexlab{}.
\newblock \bibinfo{title}{Five years ago today, we put the first quantum
  computer on the cloud. Here’s how we did it.}
\newblock
  \bibinfo{howpublished}{\url{https://research.ibm.com/blog/quantum-five-years}}.
\newblock
\newblock
\shownote{Accessed: 2022-07-27}.


\bibitem[Agler et~al\mbox{.}(2016)]%
        {agler2016microbial}
\bibfield{author}{\bibinfo{person}{Matthew~T Agler}, \bibinfo{person}{Jonas
  Ruhe}, \bibinfo{person}{Samuel Kroll}, \bibinfo{person}{Constanze Morhenn},
  \bibinfo{person}{Sang-Tae Kim}, \bibinfo{person}{Detlef Weigel}, {and}
  \bibinfo{person}{Eric~M Kemen}.} \bibinfo{year}{2016}\natexlab{}.
\newblock \showarticletitle{Microbial hub taxa link host and abiotic factors to
  plant microbiome variation}.
\newblock \bibinfo{journal}{\emph{PLoS biology}} \bibinfo{volume}{14},
  \bibinfo{number}{1} (\bibinfo{year}{2016}), \bibinfo{pages}{e1002352}.
\newblock


\bibitem[AI(2021)]%
        {sycamoredatasheet}
\bibfield{author}{\bibinfo{person}{Google~Quantum AI}.}
  \bibinfo{year}{Accessed: June 19, 2021}\natexlab{}.
\newblock \bibinfo{title}{Quantum Computer Datasheet}.
\newblock
\newblock
\newblock
\shownote{\url{https://quantumai.google/hardware/datasheet/weber.pdf}}.


\bibitem[A{\i}t-Sahalia and Matthys(2015)]%
        {ait2015robust}
\bibfield{author}{\bibinfo{person}{Yacine A{\i}t-Sahalia} {and}
  \bibinfo{person}{Felix~HA Matthys}.} \bibinfo{year}{2015}\natexlab{}.
\newblock \bibinfo{title}{Robust portfolio optimization with jumps}.
\newblock
\newblock


\bibitem[Akbarzadeh et~al\mbox{.}(2018)]%
        {akbarzadeh2018look}
\bibfield{author}{\bibinfo{person}{Meisam Akbarzadeh}, \bibinfo{person}{Soroush
  Memarmontazerin}, {and} \bibinfo{person}{Sheida Soleimani}.}
  \bibinfo{year}{2018}\natexlab{}.
\newblock \showarticletitle{Where to look for power Laws in urban road
  networks?}
\newblock \bibinfo{journal}{\emph{Applied Network Science}}
  \bibinfo{volume}{3}, \bibinfo{number}{1} (\bibinfo{year}{2018}),
  \bibinfo{pages}{1--11}.
\newblock


\bibitem[Alam et~al\mbox{.}(2020)]%
        {alam2020circuit}
\bibfield{author}{\bibinfo{person}{Mahabubul Alam}, \bibinfo{person}{Abdullah
  Ash-Saki}, {and} \bibinfo{person}{Swaroop Ghosh}.}
  \bibinfo{year}{2020}\natexlab{}.
\newblock \showarticletitle{Circuit compilation methodologies for quantum
  approximate optimization algorithm}. In \bibinfo{booktitle}{\emph{2020 53rd
  Annual IEEE/ACM International Symposium on Microarchitecture (MICRO)}}. IEEE,
  \bibinfo{pages}{215--228}.
\newblock


\bibitem[Albert(2005)]%
        {albert2005scale}
\bibfield{author}{\bibinfo{person}{Reka Albert}.}
  \bibinfo{year}{2005}\natexlab{}.
\newblock \showarticletitle{Scale-free networks in cell biology}.
\newblock \bibinfo{journal}{\emph{Journal of cell science}}
  \bibinfo{volume}{118}, \bibinfo{number}{21} (\bibinfo{year}{2005}),
  \bibinfo{pages}{4947--4957}.
\newblock


\bibitem[Amazon(2022)]%
        {AmazonBraKet}
\bibfield{author}{\bibinfo{person}{Amazon}.} \bibinfo{year}{2022}\natexlab{}.
\newblock \bibinfo{title}{{ Amazon Braket - Explore and experiment with quantum
  computing:}}.
\newblock \bibinfo{howpublished}{\url{https://aws.amazon.com/braket/}}.
\newblock
\newblock
\shownote{[Online; accessed 22-July-2021]}.


\bibitem[Arute et~al\mbox{.}(2019)]%
        {arute2019quantum}
\bibfield{author}{\bibinfo{person}{Frank Arute}, \bibinfo{person}{Kunal Arya},
  \bibinfo{person}{Ryan Babbush}, \bibinfo{person}{Dave Bacon},
  \bibinfo{person}{Joseph~C Bardin}, \bibinfo{person}{Rami Barends},
  \bibinfo{person}{Rupak Biswas}, \bibinfo{person}{Sergio Boixo},
  \bibinfo{person}{Fernando~GSL Brandao}, \bibinfo{person}{David~A Buell},
  {et~al\mbox{.}}} \bibinfo{year}{2019}\natexlab{}.
\newblock \showarticletitle{Quantum supremacy using a programmable
  superconducting processor}.
\newblock \bibinfo{journal}{\emph{Nature}} \bibinfo{volume}{574},
  \bibinfo{number}{7779} (\bibinfo{year}{2019}), \bibinfo{pages}{505--510}.
\newblock


\bibitem[Ayanzadeh et~al\mbox{.}(2021a)]%
        {ayanzadeh2021equal}
\bibfield{author}{\bibinfo{person}{Ramin Ayanzadeh}, \bibinfo{person}{Poulami
  Das}, \bibinfo{person}{Swamit~S Tannu}, {and} \bibinfo{person}{Moinuddin
  Qureshi}.} \bibinfo{year}{2021}\natexlab{a}.
\newblock \showarticletitle{EQUAL: Improving the Fidelity of Quantum Annealers
  by Injecting Controlled Perturbations}.
\newblock \bibinfo{journal}{\emph{arXiv preprint arXiv:2108.10964}}
  (\bibinfo{year}{2021}).
\newblock


\bibitem[Ayanzadeh et~al\mbox{.}(2021b)]%
        {ayanzadeh2020multi}
\bibfield{author}{\bibinfo{person}{Ramin Ayanzadeh}, \bibinfo{person}{John
  Dorband}, \bibinfo{person}{Milton Halem}, {and} \bibinfo{person}{Tim Finin}.}
  \bibinfo{year}{2021}\natexlab{b}.
\newblock \showarticletitle{Multi-Qubit Correction for Quantum Annealers}.
\newblock \bibinfo{journal}{\emph{Scientific Reports}}  \bibinfo{volume}{11}
  (\bibinfo{year}{2021}).
\newblock


\bibitem[Ayanzadeh et~al\mbox{.}(2022)]%
        {ayanzadeh2022quantum}
\bibfield{author}{\bibinfo{person}{Ramin Ayanzadeh}, \bibinfo{person}{John
  Dorband}, \bibinfo{person}{Milton Halem}, {and} \bibinfo{person}{Tim Finin}.}
  \bibinfo{year}{2022}\natexlab{}.
\newblock \showarticletitle{Quantum-assisted greedy algorithms}. In
  \bibinfo{booktitle}{\emph{IGARSS 2022-2022 IEEE International Geoscience and
  Remote Sensing Symposium}}. IEEE, \bibinfo{pages}{4911--4914}.
\newblock


\bibitem[Ayanzadeh et~al\mbox{.}(2020a)]%
        {ayanzadeh2020ensemble}
\bibfield{author}{\bibinfo{person}{Ramin Ayanzadeh}, \bibinfo{person}{Milton
  Halem}, {and} \bibinfo{person}{Tim Finin}.} \bibinfo{year}{2020}\natexlab{a}.
\newblock \showarticletitle{An ensemble approach for compressive sensing with
  quantum annealers}. In \bibinfo{booktitle}{\emph{IGARSS 2020-2020 IEEE
  International Geoscience and Remote Sensing Symposium}}. IEEE,
  \bibinfo{pages}{3517--3520}.
\newblock


\bibitem[Ayanzadeh et~al\mbox{.}(2020b)]%
        {ayanzadeh2020reinforcement}
\bibfield{author}{\bibinfo{person}{Ramin Ayanzadeh}, \bibinfo{person}{Milton
  Halem}, {and} \bibinfo{person}{Tim Finin}.} \bibinfo{year}{2020}\natexlab{b}.
\newblock \showarticletitle{Reinforcement Quantum Annealing: A Hybrid Quantum
  Learning Automata}.
\newblock \bibinfo{journal}{\emph{Scientific Reports}} \bibinfo{volume}{10},
  \bibinfo{number}{1} (\bibinfo{year}{2020}), \bibinfo{pages}{1--11}.
\newblock


\bibitem[Ayanzadeh et~al\mbox{.}(2019)]%
        {ayanzadeh2019quantum}
\bibfield{author}{\bibinfo{person}{Ramin Ayanzadeh},
  \bibinfo{person}{Seyedahmad Mousavi}, \bibinfo{person}{Milton Halem}, {and}
  \bibinfo{person}{Tim Finin}.} \bibinfo{year}{2019}\natexlab{}.
\newblock \showarticletitle{Quantum annealing based binary compressive sensing
  with matrix uncertainty}.
\newblock \bibinfo{journal}{\emph{arXiv preprint arXiv:1901.00088}}
  (\bibinfo{year}{2019}).
\newblock


\bibitem[Azad et~al\mbox{.}(2022)]%
        {9774961}
\bibfield{author}{\bibinfo{person}{Utkarsh Azad}, \bibinfo{person}{Bikash~K.
  Behera}, \bibinfo{person}{Emad~A. Ahmed}, \bibinfo{person}{Prasanta~K.
  Panigrahi}, {and} \bibinfo{person}{Ahmed Farouk}.}
  \bibinfo{year}{2022}\natexlab{}.
\newblock \showarticletitle{Solving Vehicle Routing Problem Using Quantum
  Approximate Optimization Algorithm}.
\newblock \bibinfo{journal}{\emph{IEEE Transactions on Intelligent
  Transportation Systems}} (\bibinfo{year}{2022}), \bibinfo{pages}{1--10}.
\newblock
\urldef\tempurl%
\url{https://doi.org/10.1109/TITS.2022.3172241}
\showDOI{\tempurl}


\bibitem[Baker and Radha(2022)]%
        {baker2022wasserstein}
\bibfield{author}{\bibinfo{person}{Jack~S Baker} {and}
  \bibinfo{person}{Santosh~Kumar Radha}.} \bibinfo{year}{2022}\natexlab{}.
\newblock \showarticletitle{Wasserstein Solution Quality and the Quantum
  Approximate Optimization Algorithm: A Portfolio Optimization Case Study}.
\newblock \bibinfo{journal}{\emph{arXiv preprint arXiv:2202.06782}}
  (\bibinfo{year}{2022}).
\newblock


\bibitem[Barab{\'a}si and Albert(1999)]%
        {barabasi1999emergence}
\bibfield{author}{\bibinfo{person}{Albert-L{\'a}szl{\'o} Barab{\'a}si} {and}
  \bibinfo{person}{R{\'e}ka Albert}.} \bibinfo{year}{1999}\natexlab{}.
\newblock \showarticletitle{Emergence of scaling in random networks}.
\newblock \bibinfo{journal}{\emph{science}} \bibinfo{volume}{286},
  \bibinfo{number}{5439} (\bibinfo{year}{1999}), \bibinfo{pages}{509--512}.
\newblock


\bibitem[Barab{\'a}si et~al\mbox{.}(2000)]%
        {barabasi2000scale}
\bibfield{author}{\bibinfo{person}{Albert-L{\'a}szl{\'o} Barab{\'a}si},
  \bibinfo{person}{R{\'e}ka Albert}, {and} \bibinfo{person}{Hawoong Jeong}.}
  \bibinfo{year}{2000}\natexlab{}.
\newblock \showarticletitle{Scale-free characteristics of random networks: the
  topology of the world-wide web}.
\newblock \bibinfo{journal}{\emph{Physica A: statistical mechanics and its
  applications}} \bibinfo{volume}{281}, \bibinfo{number}{1-4}
  (\bibinfo{year}{2000}), \bibinfo{pages}{69--77}.
\newblock


\bibitem[Barkoutsos et~al\mbox{.}(2020)]%
        {Barkoutsos_2020}
\bibfield{author}{\bibinfo{person}{Panagiotis~Kl. Barkoutsos},
  \bibinfo{person}{Giacomo Nannicini}, \bibinfo{person}{Anton Robert},
  \bibinfo{person}{Ivano Tavernelli}, {and} \bibinfo{person}{Stefan Woerner}.}
  \bibinfo{year}{2020}\natexlab{}.
\newblock \showarticletitle{Improving Variational Quantum Optimization using
  {CVaR}}.
\newblock \bibinfo{journal}{\emph{Quantum}}  \bibinfo{volume}{4}
  (\bibinfo{date}{apr} \bibinfo{year}{2020}), \bibinfo{pages}{256}.
\newblock
\urldef\tempurl%
\url{https://doi.org/10.22331/q-2020-04-20-256}
\showDOI{\tempurl}


\bibitem[Barron and Wood(2020)]%
        {barron2020measurement}
\bibfield{author}{\bibinfo{person}{George~S Barron} {and}
  \bibinfo{person}{Christopher~J Wood}.} \bibinfo{year}{2020}\natexlab{}.
\newblock \showarticletitle{Measurement error mitigation for variational
  quantum algorithms}.
\newblock \bibinfo{journal}{\emph{arXiv preprint arXiv:2010.08520}}
  (\bibinfo{year}{2020}).
\newblock


\bibitem[Basso et~al\mbox{.}(2021)]%
        {basso2021quantum}
\bibfield{author}{\bibinfo{person}{Joao Basso}, \bibinfo{person}{Edward Farhi},
  \bibinfo{person}{Kunal Marwaha}, \bibinfo{person}{Benjamin Villalonga}, {and}
  \bibinfo{person}{Leo Zhou}.} \bibinfo{year}{2021}\natexlab{}.
\newblock \showarticletitle{The Quantum Approximate Optimization Algorithm at
  High Depth for MaxCut on Large-Girth Regular Graphs and the
  Sherrington-Kirkpatrick Model}.
\newblock \bibinfo{journal}{\emph{arXiv preprint arXiv:2110.14206}}
  (\bibinfo{year}{2021}).
\newblock


\bibitem[Bentley et~al\mbox{.}(2022)]%
        {bentley2022quantum}
\bibfield{author}{\bibinfo{person}{Christopher~DB Bentley},
  \bibinfo{person}{Samuel Marsh}, \bibinfo{person}{Andr{\'e}~RR Carvalho},
  \bibinfo{person}{Philip Kilby}, {and} \bibinfo{person}{Michael~J Biercuk}.}
  \bibinfo{year}{2022}\natexlab{}.
\newblock \showarticletitle{Quantum computing for transport optimization}.
\newblock \bibinfo{journal}{\emph{arXiv preprint arXiv:2206.07313}}
  (\bibinfo{year}{2022}).
\newblock


\bibitem[Berliant and Watanabe(2018)]%
        {berliant2018scale}
\bibfield{author}{\bibinfo{person}{Marcus Berliant} {and}
  \bibinfo{person}{Axel~H Watanabe}.} \bibinfo{year}{2018}\natexlab{}.
\newblock \showarticletitle{A scale-free transportation network explains the
  city-size distribution}.
\newblock \bibinfo{journal}{\emph{Quantitative Economics}} \bibinfo{volume}{9},
  \bibinfo{number}{3} (\bibinfo{year}{2018}), \bibinfo{pages}{1419--1451}.
\newblock


\bibitem[Brandhofer et~al\mbox{.}(2022)]%
        {https://doi.org/10.48550/arxiv.2207.10555}
\bibfield{author}{\bibinfo{person}{Sebastian Brandhofer},
  \bibinfo{person}{Daniel Braun}, \bibinfo{person}{Vanessa Dehn},
  \bibinfo{person}{Gerhard Hellstern}, \bibinfo{person}{Matthias Hüls},
  \bibinfo{person}{Yanjun Ji}, \bibinfo{person}{Ilia Polian},
  \bibinfo{person}{Amandeep~Singh Bhatia}, {and} \bibinfo{person}{Thomas
  Wellens}.} \bibinfo{year}{2022}\natexlab{}.
\newblock \bibinfo{title}{Benchmarking the performance of portfolio
  optimization with QAOA}.
\newblock
\newblock
\urldef\tempurl%
\url{https://doi.org/10.48550/ARXIV.2207.10555}
\showDOI{\tempurl}


\bibitem[Bravyi et~al\mbox{.}(2020)]%
        {bravyi2020mitigating}
\bibfield{author}{\bibinfo{person}{Sergey Bravyi}, \bibinfo{person}{Sarah
  Sheldon}, \bibinfo{person}{Abhinav Kandala}, \bibinfo{person}{David~C Mckay},
  {and} \bibinfo{person}{Jay~M Gambetta}.} \bibinfo{year}{2020}\natexlab{}.
\newblock \showarticletitle{Mitigating measurement errors in multi-qubit
  experiments}.
\newblock \bibinfo{journal}{\emph{arXiv preprint arXiv:2006.14044}}
  (\bibinfo{year}{2020}).
\newblock


\bibitem[Bravyi et~al\mbox{.}(2016)]%
        {bravyi2016trading}
\bibfield{author}{\bibinfo{person}{Sergey Bravyi}, \bibinfo{person}{Graeme
  Smith}, {and} \bibinfo{person}{John~A Smolin}.}
  \bibinfo{year}{2016}\natexlab{}.
\newblock \showarticletitle{Trading classical and quantum computational
  resources}.
\newblock \bibinfo{journal}{\emph{Physical Review X}} \bibinfo{volume}{6},
  \bibinfo{number}{2} (\bibinfo{year}{2016}), \bibinfo{pages}{021043}.
\newblock


\bibitem[Bulancea~Lindvall(2019)]%
        {bulancea2019quantum}
\bibfield{author}{\bibinfo{person}{Oscar Bulancea~Lindvall}.}
  \bibinfo{year}{2019}\natexlab{}.
\newblock \bibinfo{title}{Quantum Methods for Sequence Alignment and
  Metagenomics}.
\newblock
\newblock


\bibitem[Buldyrev(2006)]%
        {buldyrev2006power}
\bibfield{author}{\bibinfo{person}{Sergey~V Buldyrev}.}
  \bibinfo{year}{2006}\natexlab{}.
\newblock \showarticletitle{Power law correlations in DNA sequences}.
\newblock \bibinfo{journal}{\emph{Power laws, scale-free networks and genome
  biology}} (\bibinfo{year}{2006}), \bibinfo{pages}{123--164}.
\newblock


\bibitem[Castelvecchi(2017)]%
        {castelvecchi2017ibm}
\bibfield{author}{\bibinfo{person}{Davide Castelvecchi}.}
  \bibinfo{year}{2017}\natexlab{}.
\newblock \showarticletitle{IBM's quantum cloud computer goes commercial}.
\newblock \bibinfo{journal}{\emph{Nature}} \bibinfo{volume}{543},
  \bibinfo{number}{7644} (\bibinfo{year}{2017}).
\newblock


\bibitem[Choi et~al\mbox{.}(2020)]%
        {choi2020quantum}
\bibfield{author}{\bibinfo{person}{Jaeho Choi}, \bibinfo{person}{Seunghyeok
  Oh}, {and} \bibinfo{person}{Joongheon Kim}.} \bibinfo{year}{2020}\natexlab{}.
\newblock \showarticletitle{Quantum approximation for wireless scheduling}.
\newblock \bibinfo{journal}{\emph{Applied Sciences}} \bibinfo{volume}{10},
  \bibinfo{number}{20} (\bibinfo{year}{2020}), \bibinfo{pages}{7116}.
\newblock


\bibitem[Clauset et~al\mbox{.}(2016)]%
        {clauset2016colorado}
\bibfield{author}{\bibinfo{person}{Aaron Clauset}, \bibinfo{person}{Ellen
  Tucker}, {and} \bibinfo{person}{Matthias Sainz}.}
  \bibinfo{year}{2016}\natexlab{}.
\newblock \showarticletitle{The Colorado index of complex networks}.
\newblock \bibinfo{journal}{\emph{Retrieved July}} \bibinfo{volume}{20},
  \bibinfo{number}{2018} (\bibinfo{year}{2016}), \bibinfo{pages}{22}.
\newblock


\bibitem[Cordier et~al\mbox{.}(2021)]%
        {cordier2021biology}
\bibfield{author}{\bibinfo{person}{Benjamin~A Cordier},
  \bibinfo{person}{Nicolas~PD Sawaya}, \bibinfo{person}{Gian~G Guerreschi},
  {and} \bibinfo{person}{Shannon~K McWeeney}.} \bibinfo{year}{2021}\natexlab{}.
\newblock \showarticletitle{Biology and medicine in the landscape of quantum
  advantages}.
\newblock \bibinfo{journal}{\emph{arXiv preprint arXiv:2112.00760}}
  (\bibinfo{year}{2021}).
\newblock


\bibitem[Corporation(2021)]%
        {IBMQ}
\bibfield{author}{\bibinfo{person}{International Business~Machines
  Corporation}.} \bibinfo{year}{2021}\natexlab{}.
\newblock \bibinfo{title}{{Universal Quantum Computer Development at IBM:}}.
\newblock
  \bibinfo{howpublished}{\url{http://research.ibm.com/ibm-q/research/}}.
\newblock
\newblock
\shownote{[Online; accessed 22-July-2021]}.


\bibitem[Costa et~al\mbox{.}(2019)]%
        {costa2019analysis}
\bibfield{author}{\bibinfo{person}{MO Costa}, \bibinfo{person}{R Silva},
  \bibinfo{person}{DHAL Anselmo}, {and} \bibinfo{person}{JRP Silva}.}
  \bibinfo{year}{2019}\natexlab{}.
\newblock \showarticletitle{Analysis of human DNA through power-law
  statistics}.
\newblock \bibinfo{journal}{\emph{Physical Review E}} \bibinfo{volume}{99},
  \bibinfo{number}{2} (\bibinfo{year}{2019}), \bibinfo{pages}{022112}.
\newblock


\bibitem[Dalyac et~al\mbox{.}(2021)]%
        {dalyac2021qualifying}
\bibfield{author}{\bibinfo{person}{Constantin Dalyac},
  \bibinfo{person}{Lo{\"\i}c Henriet}, \bibinfo{person}{Emmanuel Jeandel},
  \bibinfo{person}{Wolfgang Lechner}, \bibinfo{person}{Simon Perdrix},
  \bibinfo{person}{Marc Porcheron}, {and} \bibinfo{person}{Margarita
  Veshchezerova}.} \bibinfo{year}{2021}\natexlab{}.
\newblock \showarticletitle{Qualifying quantum approaches for hard industrial
  optimization problems. A case study in the field of smart-charging of
  electric vehicles}.
\newblock \bibinfo{journal}{\emph{EPJ Quantum Technology}} \bibinfo{volume}{8},
  \bibinfo{number}{1} (\bibinfo{year}{2021}), \bibinfo{pages}{12}.
\newblock


\bibitem[Das and Chakrabarti(2008)]%
        {das2008colloquium}
\bibfield{author}{\bibinfo{person}{Arnab Das} {and} \bibinfo{person}{Bikas~K
  Chakrabarti}.} \bibinfo{year}{2008}\natexlab{}.
\newblock \showarticletitle{Colloquium: Quantum annealing and analog quantum
  computation}.
\newblock \bibinfo{journal}{\emph{Reviews of Modern Physics}}
  \bibinfo{volume}{80}, \bibinfo{number}{3} (\bibinfo{year}{2008}),
  \bibinfo{pages}{1061}.
\newblock


\bibitem[Das et~al\mbox{.}(2021c)]%
        {das2021adapt}
\bibfield{author}{\bibinfo{person}{Poulami Das}, \bibinfo{person}{Swamit
  Tannu}, \bibinfo{person}{Siddharth Dangwal}, {and} \bibinfo{person}{Moinuddin
  Qureshi}.} \bibinfo{year}{2021}\natexlab{c}.
\newblock \showarticletitle{ADAPT: Mitigating Idling Errors in Qubits via
  Adaptive Dynamical Decoupling}. In \bibinfo{booktitle}{\emph{MICRO-54}}.
  \bibinfo{pages}{950--962}.
\newblock
\urldef\tempurl%
\url{https://doi.org/10.1145/3466752.3480059}
\showDOI{\tempurl}


\bibitem[Das et~al\mbox{.}(2021a)]%
        {das2021jigsaw}
\bibfield{author}{\bibinfo{person}{Poulami Das}, \bibinfo{person}{Swamit
  Tannu}, {and} \bibinfo{person}{Moinuddin Qureshi}.}
  \bibinfo{year}{2021}\natexlab{a}.
\newblock \showarticletitle{JigSaw: Boosting Fidelity of NISQ Programs via
  Measurement Subsetting}. In \bibinfo{booktitle}{\emph{MICRO-54}}.
  \bibinfo{pages}{937--949}.
\newblock
\urldef\tempurl%
\url{https://doi.org/10.1145/3466752.3480044}
\showDOI{\tempurl}


\bibitem[Das et~al\mbox{.}(2019)]%
        {das2019case}
\bibfield{author}{\bibinfo{person}{Poulami Das}, \bibinfo{person}{Swamit~S
  Tannu}, \bibinfo{person}{Prashant~J Nair}, {and} \bibinfo{person}{Moinuddin
  Qureshi}.} \bibinfo{year}{2019}\natexlab{}.
\newblock \showarticletitle{A case for multi-programming quantum computers}. In
  \bibinfo{booktitle}{\emph{Proceedings of the 52nd Annual IEEE/ACM
  International Symposium on Microarchitecture}}. \bibinfo{pages}{291--303}.
\newblock


\bibitem[Das et~al\mbox{.}(2021b)]%
        {jigsaw}
\bibfield{author}{\bibinfo{person}{Poulami Das}, \bibinfo{person}{Swamit~S
  Tannu}, {and} \bibinfo{person}{Moinuddin Qureshi}.}
  \bibinfo{year}{2021}\natexlab{b}.
\newblock \showarticletitle{JigSaw:Boosting Fidelity of NISQ Programs via
  Measurement Subsetting}. In \bibinfo{booktitle}{\emph{MICRO}}.
\newblock


\bibitem[De{\i} et~al\mbox{.}(2006)]%
        {dei2006exact}
\bibfield{author}{\bibinfo{person}{Vladimir~G De{\i}}, \bibinfo{person}{Bettina
  Klinz}, \bibinfo{person}{Gerhard~J Woeginger}, {et~al\mbox{.}}}
  \bibinfo{year}{2006}\natexlab{}.
\newblock \showarticletitle{Exact algorithms for the Hamiltonian cycle problem
  in planar graphs}.
\newblock \bibinfo{journal}{\emph{Operations Research Letters}}
  \bibinfo{volume}{34}, \bibinfo{number}{3} (\bibinfo{year}{2006}),
  \bibinfo{pages}{269--274}.
\newblock


\bibitem[Egger et~al\mbox{.}(2020)]%
        {egger2020quantum}
\bibfield{author}{\bibinfo{person}{Daniel~J Egger}, \bibinfo{person}{Claudio
  Gambella}, \bibinfo{person}{Jakub Marecek}, \bibinfo{person}{Scott McFaddin},
  \bibinfo{person}{Martin Mevissen}, \bibinfo{person}{Rudy Raymond},
  \bibinfo{person}{Andrea Simonetto}, \bibinfo{person}{Stefan Woerner}, {and}
  \bibinfo{person}{Elena Yndurain}.} \bibinfo{year}{2020}\natexlab{}.
\newblock \showarticletitle{Quantum computing for finance: State-of-the-art and
  future prospects}.
\newblock \bibinfo{journal}{\emph{IEEE Transactions on Quantum Engineering}}
  \bibinfo{volume}{1} (\bibinfo{year}{2020}), \bibinfo{pages}{1--24}.
\newblock


\bibitem[Elmerraji(2021)]%
        {elmerraji2021optimal}
\bibfield{author}{\bibinfo{person}{Jonas Elmerraji}.}
  \bibinfo{year}{2021}\natexlab{}.
\newblock \showarticletitle{Optimal CVaR Portfolio Construction Under Power Law
  Stochastic Walks}.
\newblock \bibinfo{journal}{\emph{Available at SSRN 3959932}}
  (\bibinfo{year}{2021}).
\newblock


\bibitem[Emani et~al\mbox{.}(2021)]%
        {emani2021quantum}
\bibfield{author}{\bibinfo{person}{Prashant~S Emani}, \bibinfo{person}{Jonathan
  Warrell}, \bibinfo{person}{Alan Anticevic}, \bibinfo{person}{Stefan
  Bekiranov}, \bibinfo{person}{Michael Gandal}, \bibinfo{person}{Michael~J
  McConnell}, \bibinfo{person}{Guillermo Sapiro}, \bibinfo{person}{Al{\'a}n
  Aspuru-Guzik}, \bibinfo{person}{Justin~T Baker}, \bibinfo{person}{Matteo
  Bastiani}, {et~al\mbox{.}}} \bibinfo{year}{2021}\natexlab{}.
\newblock \showarticletitle{Quantum computing at the frontiers of biological
  sciences}.
\newblock \bibinfo{journal}{\emph{Nature Methods}} \bibinfo{volume}{18},
  \bibinfo{number}{7} (\bibinfo{year}{2021}), \bibinfo{pages}{701--709}.
\newblock


\bibitem[Farhi et~al\mbox{.}(2014a)]%
        {farhi2014quantum}
\bibfield{author}{\bibinfo{person}{Edward Farhi}, \bibinfo{person}{Jeffrey
  Goldstone}, {and} \bibinfo{person}{Sam Gutmann}.}
  \bibinfo{year}{2014}\natexlab{a}.
\newblock \showarticletitle{A quantum approximate optimization algorithm}.
\newblock \bibinfo{journal}{\emph{arXiv preprint arXiv:1411.4028}}
  (\bibinfo{year}{2014}).
\newblock


\bibitem[Farhi et~al\mbox{.}(2014b)]%
        {farhi2014quantum_applied}
\bibfield{author}{\bibinfo{person}{Edward Farhi}, \bibinfo{person}{Jeffrey
  Goldstone}, {and} \bibinfo{person}{Sam Gutmann}.}
  \bibinfo{year}{2014}\natexlab{b}.
\newblock \showarticletitle{A Quantum Approximate Optimization Algorithm
  Applied to a Bounded Occurrence Constraint Problem}.
\newblock \bibinfo{journal}{\emph{arXiv preprint arXiv:1412.6062}}
  (\bibinfo{year}{2014}).
\newblock


\bibitem[Fingerhuth et~al\mbox{.}(2018)]%
        {https://doi.org/10.48550/arxiv.1810.13411}
\bibfield{author}{\bibinfo{person}{Mark Fingerhuth}, \bibinfo{person}{Tomáš
  Babej}, {and} \bibinfo{person}{Christopher Ing}.}
  \bibinfo{year}{2018}\natexlab{}.
\newblock \bibinfo{title}{A quantum alternating operator ansatz with hard and
  soft constraints for lattice protein folding}.
\newblock
\newblock
\urldef\tempurl%
\url{https://doi.org/10.48550/ARXIV.1810.13411}
\showDOI{\tempurl}


\bibitem[Fitzek et~al\mbox{.}(2021)]%
        {https://doi.org/10.48550/arxiv.2110.06799}
\bibfield{author}{\bibinfo{person}{David Fitzek}, \bibinfo{person}{Toheed
  Ghandriz}, \bibinfo{person}{Leo Laine}, \bibinfo{person}{Mats Granath}, {and}
  \bibinfo{person}{Anton~Frisk Kockum}.} \bibinfo{year}{2021}\natexlab{}.
\newblock \bibinfo{title}{Applying quantum approximate optimization to the
  heterogeneous vehicle routing problem}.
\newblock
\newblock
\urldef\tempurl%
\url{https://doi.org/10.48550/ARXIV.2110.06799}
\showDOI{\tempurl}


\bibitem[Gamermann et~al\mbox{.}(2019)]%
        {gamermann2019comprehensive}
\bibfield{author}{\bibinfo{person}{D Gamermann}, \bibinfo{person}{J
  Triana-Dopico}, {and} \bibinfo{person}{R Jaime}.}
  \bibinfo{year}{2019}\natexlab{}.
\newblock \showarticletitle{A comprehensive statistical study of metabolic and
  protein--protein interaction network properties}.
\newblock \bibinfo{journal}{\emph{Physica A: Statistical Mechanics and its
  Applications}}  \bibinfo{volume}{534} (\bibinfo{year}{2019}),
  \bibinfo{pages}{122204}.
\newblock


\bibitem[Goh et~al\mbox{.}(2002)]%
        {goh2002classification}
\bibfield{author}{\bibinfo{person}{Kwang-Il Goh}, \bibinfo{person}{Eulsik Oh},
  \bibinfo{person}{Hawoong Jeong}, \bibinfo{person}{Byungnam Kahng}, {and}
  \bibinfo{person}{Doochul Kim}.} \bibinfo{year}{2002}\natexlab{}.
\newblock \showarticletitle{Classification of scale-free networks}.
\newblock \bibinfo{journal}{\emph{Proceedings of the National Academy of
  Sciences}} \bibinfo{volume}{99}, \bibinfo{number}{20} (\bibinfo{year}{2002}),
  \bibinfo{pages}{12583--12588}.
\newblock


\bibitem[Gokhale et~al\mbox{.}(2019)]%
        {gokhale2019partial}
\bibfield{author}{\bibinfo{person}{Pranav Gokhale}, \bibinfo{person}{Yongshan
  Ding}, \bibinfo{person}{Thomas Propson}, \bibinfo{person}{Christopher
  Winkler}, \bibinfo{person}{Nelson Leung}, \bibinfo{person}{Yunong Shi},
  \bibinfo{person}{David~I Schuster}, \bibinfo{person}{Henry Hoffmann}, {and}
  \bibinfo{person}{Frederic~T Chong}.} \bibinfo{year}{2019}\natexlab{}.
\newblock \showarticletitle{{Partial Compilation of Variational Algorithms for
  Noisy Intermediate-Scale Quantum Machines}}. In
  \bibinfo{booktitle}{\emph{Proceedings of the 52nd Annual IEEE/ACM
  International Symposium on Microarchitecture}}. ACM,
  \bibinfo{pages}{266--278}.
\newblock


\bibitem[Gokhale et~al\mbox{.}(2020)]%
        {gokhale2020optimized}
\bibfield{author}{\bibinfo{person}{Pranav Gokhale}, \bibinfo{person}{Ali
  Javadi-Abhari}, \bibinfo{person}{Nathan Earnest}, \bibinfo{person}{Yunong
  Shi}, {and} \bibinfo{person}{Frederic~T Chong}.}
  \bibinfo{year}{2020}\natexlab{}.
\newblock \showarticletitle{{Optimized Quantum Compilation for Near-Term
  Algorithms with OpenPulse}}.
\newblock \bibinfo{journal}{\emph{arXiv preprint arXiv:2004.11205}}
  (\bibinfo{year}{2020}).
\newblock


\bibitem[Gray et~al\mbox{.}(2018)]%
        {gray2018super}
\bibfield{author}{\bibinfo{person}{Caitlin Gray}, \bibinfo{person}{Lewis
  Mitchell}, {and} \bibinfo{person}{Matthew Roughan}.}
  \bibinfo{year}{2018}\natexlab{}.
\newblock \showarticletitle{Super-blockers and the effect of network structure
  on information cascades}. In \bibinfo{booktitle}{\emph{Companion Proceedings
  of the The Web Conference 2018}}. \bibinfo{pages}{1435--1441}.
\newblock


\bibitem[Guerreschi and Matsuura(2019)]%
        {guerreschi2019qaoa}
\bibfield{author}{\bibinfo{person}{Gian~Giacomo Guerreschi} {and}
  \bibinfo{person}{Anne~Y Matsuura}.} \bibinfo{year}{2019}\natexlab{}.
\newblock \showarticletitle{QAOA for Max-Cut requires hundreds of qubits for
  quantum speed-up}.
\newblock \bibinfo{journal}{\emph{Scientific reports}} \bibinfo{volume}{9},
  \bibinfo{number}{1} (\bibinfo{year}{2019}), \bibinfo{pages}{1--7}.
\newblock


\bibitem[Hadlock(1975)]%
        {hadlock1975finding}
\bibfield{author}{\bibinfo{person}{Frank Hadlock}.}
  \bibinfo{year}{1975}\natexlab{}.
\newblock \showarticletitle{Finding a maximum cut of a planar graph in
  polynomial time}.
\newblock \bibinfo{journal}{\emph{SIAM J. Comput.}} \bibinfo{volume}{4},
  \bibinfo{number}{3} (\bibinfo{year}{1975}), \bibinfo{pages}{221--225}.
\newblock


\bibitem[Harrigan et~al\mbox{.}(2021)]%
        {harrigan2021quantum}
\bibfield{author}{\bibinfo{person}{Matthew~P Harrigan},
  \bibinfo{person}{Kevin~J Sung}, \bibinfo{person}{Matthew Neeley},
  \bibinfo{person}{Kevin~J Satzinger}, \bibinfo{person}{Frank Arute},
  \bibinfo{person}{Kunal Arya}, \bibinfo{person}{Juan Atalaya},
  \bibinfo{person}{Joseph~C Bardin}, \bibinfo{person}{Rami Barends},
  \bibinfo{person}{Sergio Boixo}, {et~al\mbox{.}}}
  \bibinfo{year}{2021}\natexlab{}.
\newblock \showarticletitle{Quantum approximate optimization of non-planar
  graph problems on a planar superconducting processor}.
\newblock \bibinfo{journal}{\emph{Nature Physics}} \bibinfo{volume}{17},
  \bibinfo{number}{3} (\bibinfo{year}{2021}), \bibinfo{pages}{332--336}.
\newblock


\bibitem[Herrman et~al\mbox{.}(2022)]%
        {herrman2022multi}
\bibfield{author}{\bibinfo{person}{Rebekah Herrman}, \bibinfo{person}{Phillip~C
  Lotshaw}, \bibinfo{person}{James Ostrowski}, \bibinfo{person}{Travis~S
  Humble}, {and} \bibinfo{person}{George Siopsis}.}
  \bibinfo{year}{2022}\natexlab{}.
\newblock \showarticletitle{Multi-angle quantum approximate optimization
  algorithm}.
\newblock \bibinfo{journal}{\emph{Scientific Reports}} \bibinfo{volume}{12},
  \bibinfo{number}{1} (\bibinfo{year}{2022}), \bibinfo{pages}{1--10}.
\newblock


\bibitem[Hou et~al\mbox{.}(2018)]%
        {hou2018does}
\bibfield{author}{\bibinfo{person}{Yunzhang Hou}, \bibinfo{person}{Xiaoling
  Wang}, \bibinfo{person}{Yenchun~Jim Wu}, {and} \bibinfo{person}{Peixu He}.}
  \bibinfo{year}{2018}\natexlab{}.
\newblock \showarticletitle{How does the trust affect the topology of supply
  chain network and its resilience? An agent-based approach}.
\newblock \bibinfo{journal}{\emph{Transportation Research Part E: Logistics and
  Transportation Review}}  \bibinfo{volume}{116} (\bibinfo{year}{2018}),
  \bibinfo{pages}{229--241}.
\newblock


\bibitem[House et~al\mbox{.}(2015)]%
        {house2015testing}
\bibfield{author}{\bibinfo{person}{Thomas House}, \bibinfo{person}{Jonathan~M
  Read}, \bibinfo{person}{Leon Danon}, {and} \bibinfo{person}{Matthew~J
  Keeling}.} \bibinfo{year}{2015}\natexlab{}.
\newblock \showarticletitle{Testing the hypothesis of preferential attachment
  in social network formation}.
\newblock \bibinfo{journal}{\emph{EPJ Data Science}} \bibinfo{volume}{4},
  \bibinfo{number}{1} (\bibinfo{year}{2015}), \bibinfo{pages}{1--13}.
\newblock


\bibitem[IBM(2010)]%
        {matrixmeasurementmitigation}
\bibfield{author}{\bibinfo{person}{IBM}.} \bibinfo{year}{2010}\natexlab{}.
\newblock \bibinfo{title}{{Measurement Error Mitigation}}.
\newblock
  \bibinfo{howpublished}{\url{https://qiskit.org/textbook/ch-quantum-hardware/measurement-error-mitigation.html}}.
\newblock
\newblock
\shownote{[Online; accessed 26-July-2020]}.


\bibitem[Jeong et~al\mbox{.}(2001)]%
        {jeong2001lethality}
\bibfield{author}{\bibinfo{person}{Hawoong Jeong}, \bibinfo{person}{Sean~P
  Mason}, \bibinfo{person}{A-L Barab{\'a}si}, {and} \bibinfo{person}{Zoltan~N
  Oltvai}.} \bibinfo{year}{2001}\natexlab{}.
\newblock \showarticletitle{Lethality and centrality in protein networks}.
\newblock \bibinfo{journal}{\emph{Nature}} \bibinfo{volume}{411},
  \bibinfo{number}{6833} (\bibinfo{year}{2001}), \bibinfo{pages}{41--42}.
\newblock


\bibitem[Kahng et~al\mbox{.}(2004)]%
        {kahng2004emergence}
\bibfield{author}{\bibinfo{person}{B Kahng}, \bibinfo{person}{I Yang},
  \bibinfo{person}{H Jeong}, {and} \bibinfo{person}{A-L Barab{\'a}si}.}
  \bibinfo{year}{2004}\natexlab{}.
\newblock \showarticletitle{Emergence of power-law behaviors in online
  auctions}.
\newblock In \bibinfo{booktitle}{\emph{The Application of Econophysics}}.
  \bibinfo{publisher}{Springer}, \bibinfo{pages}{204--209}.
\newblock


\bibitem[Kim et~al\mbox{.}(2022)]%
        {kim2022sparsity}
\bibfield{author}{\bibinfo{person}{Sung-Soo Kim}, \bibinfo{person}{Young-Min
  Kang}, {and} \bibinfo{person}{Young-Kuk Kimt}.}
  \bibinfo{year}{2022}\natexlab{}.
\newblock \showarticletitle{Sparsity-Aware Reachability Computation for Massive
  Graphs}. In \bibinfo{booktitle}{\emph{2022 IEEE International Conference on
  Big Data and Smart Computing (BigComp)}}. IEEE, \bibinfo{pages}{157--160}.
\newblock


\bibitem[Kwon and Bae(2020)]%
        {kwon2020hybrid}
\bibfield{author}{\bibinfo{person}{Hyeokjea Kwon} {and}
  \bibinfo{person}{Joonwoo Bae}.} \bibinfo{year}{2020}\natexlab{}.
\newblock \showarticletitle{A hybrid quantum-classical approach to mitigating
  measurement errors}.
\newblock \bibinfo{journal}{\emph{arXiv preprint arXiv:2003.12314}}
  (\bibinfo{year}{2020}).
\newblock


\bibitem[Lao and Browne(2022)]%
        {lao20222qan}
\bibfield{author}{\bibinfo{person}{Lingling Lao} {and} \bibinfo{person}{Dan~E
  Browne}.} \bibinfo{year}{2022}\natexlab{}.
\newblock \showarticletitle{2qan: A quantum compiler for 2-local qubit
  hamiltonian simulation algorithms}. In \bibinfo{booktitle}{\emph{Proceedings
  of the 49th Annual International Symposium on Computer Architecture}}.
  \bibinfo{pages}{351--365}.
\newblock


\bibitem[Li et~al\mbox{.}(2018)]%
        {li2018tackling}
\bibfield{author}{\bibinfo{person}{Gushu Li}, \bibinfo{person}{Yufei Ding},
  {and} \bibinfo{person}{Yuan Xie}.} \bibinfo{year}{2018}\natexlab{}.
\newblock \showarticletitle{{Tackling the Qubit Mapping Problem for NISQ-Era
  Quantum Devices}}.
\newblock \bibinfo{journal}{\emph{arXiv preprint arXiv:1809.02573}}
  (\bibinfo{year}{2018}).
\newblock


\bibitem[Li et~al\mbox{.}(2022)]%
        {li2022paulihedral}
\bibfield{author}{\bibinfo{person}{Gushu Li}, \bibinfo{person}{Anbang Wu},
  \bibinfo{person}{Yunong Shi}, \bibinfo{person}{Ali Javadi-Abhari},
  \bibinfo{person}{Yufei Ding}, {and} \bibinfo{person}{Yuan Xie}.}
  \bibinfo{year}{2022}\natexlab{}.
\newblock \showarticletitle{Paulihedral: a generalized block-wise compiler
  optimization framework for Quantum simulation kernels}. In
  \bibinfo{booktitle}{\emph{Proceedings of the 27th ACM International
  Conference on Architectural Support for Programming Languages and Operating
  Systems}}. \bibinfo{pages}{554--569}.
\newblock


\bibitem[Li et~al\mbox{.}(2021)]%
        {li2021large}
\bibfield{author}{\bibinfo{person}{Junde Li}, \bibinfo{person}{Mahabubul Alam},
  {and} \bibinfo{person}{Swaroop Ghosh}.} \bibinfo{year}{2021}\natexlab{}.
\newblock \showarticletitle{Large-scale Quantum Approximate Optimization via
  Divide-and-Conquer}.
\newblock \bibinfo{journal}{\emph{arXiv preprint arXiv:2102.13288}}
  (\bibinfo{year}{2021}).
\newblock


\bibitem[Liu and Dou(2021)]%
        {9407180}
\bibfield{author}{\bibinfo{person}{Lei Liu} {and} \bibinfo{person}{Xinglei
  Dou}.} \bibinfo{year}{2021}\natexlab{}.
\newblock \showarticletitle{QuCloud: A New Qubit Mapping Mechanism for
  Multi-programming Quantum Computing in Cloud Environment}. In
  \bibinfo{booktitle}{\emph{2021 IEEE International Symposium on
  High-Performance Computer Architecture (HPCA)}}. \bibinfo{pages}{167--178}.
\newblock
\urldef\tempurl%
\url{https://doi.org/10.1109/HPCA51647.2021.00024}
\showDOI{\tempurl}


\bibitem[Lloyd(2018)]%
        {lloyd2018quantum}
\bibfield{author}{\bibinfo{person}{Seth Lloyd}.}
  \bibinfo{year}{2018}\natexlab{}.
\newblock \showarticletitle{Quantum approximate optimization is computationally
  universal}.
\newblock \bibinfo{journal}{\emph{arXiv preprint arXiv:1812.11075}}
  (\bibinfo{year}{2018}).
\newblock


\bibitem[Lucas(2014)]%
        {lucas2014ising}
\bibfield{author}{\bibinfo{person}{Andrew Lucas}.}
  \bibinfo{year}{2014}\natexlab{}.
\newblock \showarticletitle{Ising formulations of many NP problems}.
\newblock \bibinfo{journal}{\emph{Frontiers in physics}}  \bibinfo{volume}{2}
  (\bibinfo{year}{2014}), \bibinfo{pages}{5}.
\newblock


\bibitem[Lusseau(2003)]%
        {lusseau2003emergent}
\bibfield{author}{\bibinfo{person}{David Lusseau}.}
  \bibinfo{year}{2003}\natexlab{}.
\newblock \showarticletitle{The emergent properties of a dolphin social
  network}.
\newblock \bibinfo{journal}{\emph{Proceedings of the Royal Society of London.
  Series B: Biological Sciences}} \bibinfo{volume}{270},
  \bibinfo{number}{suppl\_2} (\bibinfo{year}{2003}),
  \bibinfo{pages}{S186--S188}.
\newblock


\bibitem[Magner et~al\mbox{.}(2015)]%
        {magner2015origin}
\bibfield{author}{\bibinfo{person}{Abram Magner}, \bibinfo{person}{Wojciech
  Szpankowski}, {and} \bibinfo{person}{Daisuke Kihara}.}
  \bibinfo{year}{2015}\natexlab{}.
\newblock \showarticletitle{On the origin of protein superfamilies and
  superfolds}.
\newblock \bibinfo{journal}{\emph{Scientific reports}} \bibinfo{volume}{5},
  \bibinfo{number}{1} (\bibinfo{year}{2015}), \bibinfo{pages}{1--7}.
\newblock


\bibitem[Microsoft(2022)]%
        {MicrosoftAzure}
\bibfield{author}{\bibinfo{person}{Microsoft}.}
  \bibinfo{year}{2022}\natexlab{}.
\newblock \bibinfo{title}{{Azure Quantum - Quantum Service | Microsoft Azure}}.
\newblock
  \bibinfo{howpublished}{\url{https://azure.microsoft.com/en-us/services/quantum/\#product-overview}}.
\newblock
\newblock
\shownote{[Online; accessed 22-July-2021]}.


\bibitem[Mislove et~al\mbox{.}(2007)]%
        {mislove2007measurement}
\bibfield{author}{\bibinfo{person}{Alan Mislove}, \bibinfo{person}{Massimiliano
  Marcon}, \bibinfo{person}{Krishna~P Gummadi}, \bibinfo{person}{Peter
  Druschel}, {and} \bibinfo{person}{Bobby Bhattacharjee}.}
  \bibinfo{year}{2007}\natexlab{}.
\newblock \showarticletitle{Measurement and analysis of online social
  networks}. In \bibinfo{booktitle}{\emph{Proceedings of the 7th ACM SIGCOMM
  conference on Internet measurement}}. \bibinfo{pages}{29--42}.
\newblock


\bibitem[Morales et~al\mbox{.}(2020)]%
        {morales2020universality}
\bibfield{author}{\bibinfo{person}{Mauro~ES Morales}, \bibinfo{person}{Jacob~D
  Biamonte}, {and} \bibinfo{person}{Zolt{\'a}n Zimbor{\'a}s}.}
  \bibinfo{year}{2020}\natexlab{}.
\newblock \showarticletitle{On the universality of the quantum approximate
  optimization algorithm}.
\newblock \bibinfo{journal}{\emph{Quantum Information Processing}}
  \bibinfo{volume}{19}, \bibinfo{number}{9} (\bibinfo{year}{2020}),
  \bibinfo{pages}{1--26}.
\newblock


\bibitem[Mori et~al\mbox{.}(2020)]%
        {mori2020common}
\bibfield{author}{\bibinfo{person}{Tomoya Mori}, \bibinfo{person}{Tony~E
  Smith}, {and} \bibinfo{person}{Wen-Tai Hsu}.}
  \bibinfo{year}{2020}\natexlab{}.
\newblock \showarticletitle{Common power laws for cities and spatial fractal
  structures}.
\newblock \bibinfo{journal}{\emph{Proceedings of the National Academy of
  Sciences}} \bibinfo{volume}{117}, \bibinfo{number}{12}
  (\bibinfo{year}{2020}), \bibinfo{pages}{6469--6475}.
\newblock


\bibitem[Murali et~al\mbox{.}(2019)]%
        {noiseadaptive}
\bibfield{author}{\bibinfo{person}{Prakash Murali}, \bibinfo{person}{Jonathan~M
  Baker}, \bibinfo{person}{Ali~Javadi Abhari}, \bibinfo{person}{Frederic~T
  Chong}, {and} \bibinfo{person}{Margaret Martonosi}.}
  \bibinfo{year}{2019}\natexlab{}.
\newblock \showarticletitle{Noise-Adaptive Compiler Mappings for Noisy
  Intermediate-Scale Quantum Computers}.
\newblock \bibinfo{journal}{\emph{arXiv preprint arXiv:1901.11054}}
  (\bibinfo{year}{2019}).
\newblock


\bibitem[Murali et~al\mbox{.}(2020)]%
        {murali2020software}
\bibfield{author}{\bibinfo{person}{Prakash Murali}, \bibinfo{person}{David~C
  McKay}, \bibinfo{person}{Margaret Martonosi}, {and} \bibinfo{person}{Ali
  Javadi-Abhari}.} \bibinfo{year}{2020}\natexlab{}.
\newblock \showarticletitle{{Software Mitigation of Crosstalk on Noisy
  Intermediate-Scale Quantum Computers}}.
\newblock \bibinfo{journal}{\emph{arXiv preprint arXiv:2001.02826}}
  (\bibinfo{year}{2020}).
\newblock


\bibitem[Nam et~al\mbox{.}(2020)]%
        {nam2020ground}
\bibfield{author}{\bibinfo{person}{Yunseong Nam}, \bibinfo{person}{Jwo-Sy
  Chen}, \bibinfo{person}{Neal~C Pisenti}, \bibinfo{person}{Kenneth Wright},
  \bibinfo{person}{Conor Delaney}, \bibinfo{person}{Dmitri Maslov},
  \bibinfo{person}{Kenneth~R Brown}, \bibinfo{person}{Stewart Allen},
  \bibinfo{person}{Jason~M Amini}, \bibinfo{person}{Joel Apisdorf},
  {et~al\mbox{.}}} \bibinfo{year}{2020}\natexlab{}.
\newblock \showarticletitle{Ground-state energy estimation of the water
  molecule on a trapped-ion quantum computer}.
\newblock \bibinfo{journal}{\emph{npj Quantum Information}}
  \bibinfo{volume}{6}, \bibinfo{number}{1} (\bibinfo{year}{2020}),
  \bibinfo{pages}{1--6}.
\newblock


\bibitem[Nishio et~al\mbox{.}(2019)]%
        {nishio}
\bibfield{author}{\bibinfo{person}{Shin Nishio}, \bibinfo{person}{Yulu Pan},
  \bibinfo{person}{Takahiko Satoh}, \bibinfo{person}{Hideharu Amano}, {and}
  \bibinfo{person}{Rodney Van~Meter}.} \bibinfo{year}{2019}\natexlab{}.
\newblock \showarticletitle{Extracting Success from IBM's 20-Qubit Machines
  Using Error-Aware Compilation}.
\newblock \bibinfo{journal}{\emph{arXiv preprint arXiv:1903.10963}}
  (\bibinfo{year}{2019}).
\newblock


\bibitem[O'Malley et~al\mbox{.}(2018)]%
        {o2018nonnegative}
\bibfield{author}{\bibinfo{person}{Daniel O'Malley}, \bibinfo{person}{Velimir~V
  Vesselinov}, \bibinfo{person}{Boian~S Alexandrov}, {and}
  \bibinfo{person}{Ludmil~B Alexandrov}.} \bibinfo{year}{2018}\natexlab{}.
\newblock \showarticletitle{Nonnegative/binary matrix factorization with a
  D-Wave quantum annealer}.
\newblock \bibinfo{journal}{\emph{PloS one}} \bibinfo{volume}{13},
  \bibinfo{number}{12} (\bibinfo{year}{2018}), \bibinfo{pages}{e0206653}.
\newblock


\bibitem[Pastor-Satorras et~al\mbox{.}(2015)]%
        {pastor2015epidemic}
\bibfield{author}{\bibinfo{person}{Romualdo Pastor-Satorras},
  \bibinfo{person}{Claudio Castellano}, \bibinfo{person}{Piet Van~Mieghem},
  {and} \bibinfo{person}{Alessandro Vespignani}.}
  \bibinfo{year}{2015}\natexlab{}.
\newblock \showarticletitle{Epidemic processes in complex networks}.
\newblock \bibinfo{journal}{\emph{Reviews of modern physics}}
  \bibinfo{volume}{87}, \bibinfo{number}{3} (\bibinfo{year}{2015}),
  \bibinfo{pages}{925}.
\newblock


\bibitem[Patel et~al\mbox{.}(2022a)]%
        {patel2022geyser}
\bibfield{author}{\bibinfo{person}{Tirthak Patel}, \bibinfo{person}{Daniel
  Silver}, {and} \bibinfo{person}{Devesh Tiwari}.}
  \bibinfo{year}{2022}\natexlab{a}.
\newblock \showarticletitle{Geyser: a compilation framework for quantum
  computing with neutral atoms}. In \bibinfo{booktitle}{\emph{Proceedings of
  the 49th Annual International Symposium on Computer Architecture}}.
  \bibinfo{pages}{383--395}.
\newblock


\bibitem[Patel and Tiwari(2020)]%
        {patel2020veritas}
\bibfield{author}{\bibinfo{person}{Tirthak Patel} {and} \bibinfo{person}{Devesh
  Tiwari}.} \bibinfo{year}{2020}\natexlab{}.
\newblock \showarticletitle{Veritas: accurately estimating the correct output
  on noisy intermediate-scale quantum computers}. In
  \bibinfo{booktitle}{\emph{SC20: International Conference for High Performance
  Computing, Networking, Storage and Analysis}}. IEEE, \bibinfo{pages}{1--16}.
\newblock


\bibitem[Patel et~al\mbox{.}(2022b)]%
        {patel2022quest}
\bibfield{author}{\bibinfo{person}{Tirthak Patel}, \bibinfo{person}{Ed Younis},
  \bibinfo{person}{Costin Iancu}, \bibinfo{person}{Wibe de Jong}, {and}
  \bibinfo{person}{Devesh Tiwari}.} \bibinfo{year}{2022}\natexlab{b}.
\newblock \showarticletitle{QUEST: systematically approximating Quantum
  circuits for higher output fidelity}. In
  \bibinfo{booktitle}{\emph{Proceedings of the 27th ACM International
  Conference on Architectural Support for Programming Languages and Operating
  Systems}}. \bibinfo{pages}{514--528}.
\newblock


\bibitem[Peng et~al\mbox{.}(1995)]%
        {peng1995statistical}
\bibfield{author}{\bibinfo{person}{C-K Peng}, \bibinfo{person}{SV Buldyrev},
  \bibinfo{person}{AL Goldberger}, \bibinfo{person}{S Havlin},
  \bibinfo{person}{RN Mantegna}, \bibinfo{person}{M Simons}, {and}
  \bibinfo{person}{HE Stanley}.} \bibinfo{year}{1995}\natexlab{}.
\newblock \showarticletitle{Statistical properties of DNA sequences}.
\newblock \bibinfo{journal}{\emph{Physica A: Statistical Mechanics and its
  Applications}} \bibinfo{volume}{221}, \bibinfo{number}{1-3}
  (\bibinfo{year}{1995}), \bibinfo{pages}{180--192}.
\newblock


\bibitem[Peng et~al\mbox{.}(2020)]%
        {peng2020simulating}
\bibfield{author}{\bibinfo{person}{Tianyi Peng}, \bibinfo{person}{Aram~W
  Harrow}, \bibinfo{person}{Maris Ozols}, {and} \bibinfo{person}{Xiaodi Wu}.}
  \bibinfo{year}{2020}\natexlab{}.
\newblock \showarticletitle{Simulating large quantum circuits on a small
  quantum computer}.
\newblock \bibinfo{journal}{\emph{Physical Review Letters}}
  \bibinfo{volume}{125}, \bibinfo{number}{15} (\bibinfo{year}{2020}),
  \bibinfo{pages}{150504}.
\newblock


\bibitem[Preskill(2018)]%
        {preskillNISQ}
\bibfield{author}{\bibinfo{person}{John Preskill}.}
  \bibinfo{year}{2018}\natexlab{}.
\newblock \showarticletitle{Quantum Computing in the NISQ era and beyond}.
\newblock \bibinfo{journal}{\emph{arXiv preprint arXiv:1801.00862}}
  (\bibinfo{year}{2018}).
\newblock


\bibitem[Qian et~al\mbox{.}(2001)]%
        {qian2001protein}
\bibfield{author}{\bibinfo{person}{Jiang Qian}, \bibinfo{person}{Nicholas~M
  Luscombe}, {and} \bibinfo{person}{Mark Gerstein}.}
  \bibinfo{year}{2001}\natexlab{}.
\newblock \showarticletitle{Protein family and fold occurrence in genomes:
  power-law behaviour and evolutionary model}.
\newblock \bibinfo{journal}{\emph{Journal of molecular biology}}
  \bibinfo{volume}{313}, \bibinfo{number}{4} (\bibinfo{year}{2001}),
  \bibinfo{pages}{673--681}.
\newblock


\bibitem[Ravi et~al\mbox{.}(2021a)]%
        {ravi2021quantum}
\bibfield{author}{\bibinfo{person}{Gokul~Subramanian Ravi},
  \bibinfo{person}{Kaitlin~N Smith}, \bibinfo{person}{Pranav Gokhale}, {and}
  \bibinfo{person}{Frederic~T Chong}.} \bibinfo{year}{2021}\natexlab{a}.
\newblock \showarticletitle{Quantum Computing in the Cloud: Analyzing job and
  machine characteristics}. In \bibinfo{booktitle}{\emph{2021 IEEE
  International Symposium on Workload Characterization (IISWC)}}. IEEE,
  \bibinfo{pages}{39--50}.
\newblock


\bibitem[Ravi et~al\mbox{.}(2021b)]%
        {ravi2021adaptive}
\bibfield{author}{\bibinfo{person}{Gokul~Subramanian Ravi},
  \bibinfo{person}{Kaitlin~N Smith}, \bibinfo{person}{Prakash Murali}, {and}
  \bibinfo{person}{Frederic~T Chong}.} \bibinfo{year}{2021}\natexlab{b}.
\newblock \showarticletitle{Adaptive job and resource management for the
  growing quantum cloud}. In \bibinfo{booktitle}{\emph{2021 IEEE International
  Conference on Quantum Computing and Engineering (QCE)}}. IEEE,
  \bibinfo{pages}{301--312}.
\newblock


\bibitem[Rieffel et~al\mbox{.}(2015)]%
        {rieffel2015case}
\bibfield{author}{\bibinfo{person}{Eleanor~G Rieffel}, \bibinfo{person}{Davide
  Venturelli}, \bibinfo{person}{Bryan O'Gorman}, \bibinfo{person}{Minh~B Do},
  \bibinfo{person}{Elicia~M Prystay}, {and} \bibinfo{person}{Vadim~N
  Smelyanskiy}.} \bibinfo{year}{2015}\natexlab{}.
\newblock \showarticletitle{A case study in programming a quantum annealer for
  hard operational planning problems}.
\newblock \bibinfo{journal}{\emph{Quantum Information Processing}}
  \bibinfo{volume}{14}, \bibinfo{number}{1} (\bibinfo{year}{2015}),
  \bibinfo{pages}{1--36}.
\newblock


\bibitem[Robert et~al\mbox{.}(2021)]%
        {robert2021resource}
\bibfield{author}{\bibinfo{person}{Anton Robert},
  \bibinfo{person}{Panagiotis~Kl Barkoutsos}, \bibinfo{person}{Stefan Woerner},
  {and} \bibinfo{person}{Ivano Tavernelli}.} \bibinfo{year}{2021}\natexlab{}.
\newblock \showarticletitle{Resource-efficient quantum algorithm for protein
  folding}.
\newblock \bibinfo{journal}{\emph{npj Quantum Information}}
  \bibinfo{volume}{7}, \bibinfo{number}{1} (\bibinfo{year}{2021}),
  \bibinfo{pages}{1--5}.
\newblock


\bibitem[Sarkar et~al\mbox{.}(2021)]%
        {sarkar2021quaser}
\bibfield{author}{\bibinfo{person}{Aritra Sarkar}, \bibinfo{person}{Zaid
  Al-Ars}, {and} \bibinfo{person}{Koen Bertels}.}
  \bibinfo{year}{2021}\natexlab{}.
\newblock \showarticletitle{QuASeR: Quantum Accelerated de novo DNA sequence
  reconstruction}.
\newblock \bibinfo{journal}{\emph{Plos one}} \bibinfo{volume}{16},
  \bibinfo{number}{4} (\bibinfo{year}{2021}), \bibinfo{pages}{e0249850}.
\newblock


\bibitem[Sawada and Honda(2006)]%
        {sawada2006structural}
\bibfield{author}{\bibinfo{person}{Yoshito Sawada} {and}
  \bibinfo{person}{Shinya Honda}.} \bibinfo{year}{2006}\natexlab{}.
\newblock \showarticletitle{Structural diversity of protein segments follows a
  power-law distribution}.
\newblock \bibinfo{journal}{\emph{Biophysical journal}} \bibinfo{volume}{91},
  \bibinfo{number}{4} (\bibinfo{year}{2006}), \bibinfo{pages}{1213--1223}.
\newblock


\bibitem[Sherrington and Kirkpatrick(1975)]%
        {sherrington1975solvable}
\bibfield{author}{\bibinfo{person}{David Sherrington} {and}
  \bibinfo{person}{Scott Kirkpatrick}.} \bibinfo{year}{1975}\natexlab{}.
\newblock \showarticletitle{Solvable model of a spin-glass}.
\newblock \bibinfo{journal}{\emph{Physical review letters}}
  \bibinfo{volume}{35}, \bibinfo{number}{26} (\bibinfo{year}{1975}),
  \bibinfo{pages}{1792}.
\newblock


\bibitem[Shi et~al\mbox{.}(2019)]%
        {shi2019optimized}
\bibfield{author}{\bibinfo{person}{Yunong Shi}, \bibinfo{person}{Nelson Leung},
  \bibinfo{person}{Pranav Gokhale}, \bibinfo{person}{Zane Rossi},
  \bibinfo{person}{David~I Schuster}, \bibinfo{person}{Henry Hoffmann}, {and}
  \bibinfo{person}{Frederic~T Chong}.} \bibinfo{year}{2019}\natexlab{}.
\newblock \showarticletitle{Optimized compilation of aggregated instructions
  for realistic quantum computers}. In \bibinfo{booktitle}{\emph{Proceedings of
  the Twenty-Fourth International Conference on Architectural Support for
  Programming Languages and Operating Systems}}. \bibinfo{pages}{1031--1044}.
\newblock


\bibitem[Smith et~al\mbox{.}(2021)]%
        {smith2021error}
\bibfield{author}{\bibinfo{person}{Kaitlin~N Smith},
  \bibinfo{person}{Gokul~Subramanian Ravi}, \bibinfo{person}{Prakash Murali},
  \bibinfo{person}{Jonathan~M Baker}, \bibinfo{person}{Nathan Earnest},
  \bibinfo{person}{Ali Javadi-Abhari}, {and} \bibinfo{person}{Frederic~T
  Chong}.} \bibinfo{year}{2021}\natexlab{}.
\newblock \showarticletitle{Error Mitigation in Quantum Computers through
  Instruction Scheduling}.
\newblock \bibinfo{journal}{\emph{arXiv preprint arXiv:2105.01760}}
  (\bibinfo{year}{2021}).
\newblock


\bibitem[Stein et~al\mbox{.}(2022)]%
        {stein2022eqc}
\bibfield{author}{\bibinfo{person}{Samuel Stein}, \bibinfo{person}{Nathan
  Wiebe}, \bibinfo{person}{Yufei Ding}, \bibinfo{person}{Peng Bo},
  \bibinfo{person}{Karol Kowalski}, \bibinfo{person}{Nathan Baker},
  \bibinfo{person}{James Ang}, {and} \bibinfo{person}{Ang Li}.}
  \bibinfo{year}{2022}\natexlab{}.
\newblock \showarticletitle{EQC: ensembled quantum computing for variational
  quantum algorithms}. In \bibinfo{booktitle}{\emph{Proceedings of the 49th
  Annual International Symposium on Computer Architecture}}.
  \bibinfo{pages}{59--71}.
\newblock


\bibitem[Streif et~al\mbox{.}(2021)]%
        {streif2021beating}
\bibfield{author}{\bibinfo{person}{Michael Streif}, \bibinfo{person}{Sheir
  Yarkoni}, \bibinfo{person}{Andrea Skolik}, \bibinfo{person}{Florian Neukart},
  {and} \bibinfo{person}{Martin Leib}.} \bibinfo{year}{2021}\natexlab{}.
\newblock \showarticletitle{Beating classical heuristics for the binary paint
  shop problem with the quantum approximate optimization algorithm}.
\newblock \bibinfo{journal}{\emph{Physical Review A}} \bibinfo{volume}{104},
  \bibinfo{number}{1} (\bibinfo{year}{2021}), \bibinfo{pages}{012403}.
\newblock


\bibitem[Sung et~al\mbox{.}(2020)]%
        {sung2020using}
\bibfield{author}{\bibinfo{person}{Kevin~J Sung}, \bibinfo{person}{Jiahao Yao},
  \bibinfo{person}{Matthew~P Harrigan}, \bibinfo{person}{Nicholas~C Rubin},
  \bibinfo{person}{Zhang Jiang}, \bibinfo{person}{Lin Lin},
  \bibinfo{person}{Ryan Babbush}, {and} \bibinfo{person}{Jarrod~R McClean}.}
  \bibinfo{year}{2020}\natexlab{}.
\newblock \showarticletitle{Using models to improve optimizers for variational
  quantum algorithms}.
\newblock \bibinfo{journal}{\emph{Quantum Science and Technology}}
  \bibinfo{volume}{5}, \bibinfo{number}{4} (\bibinfo{year}{2020}),
  \bibinfo{pages}{044008}.
\newblock


\bibitem[Suo et~al\mbox{.}(2018)]%
        {suo2018exploring}
\bibfield{author}{\bibinfo{person}{Qi Suo}, \bibinfo{person}{Jin-Li Guo},
  \bibinfo{person}{Shiwei Sun}, {and} \bibinfo{person}{Han Liu}.}
  \bibinfo{year}{2018}\natexlab{}.
\newblock \showarticletitle{Exploring the evolutionary mechanism of complex
  supply chain systems using evolving hypergraphs}.
\newblock \bibinfo{journal}{\emph{Physica A: Statistical Mechanics and its
  Applications}}  \bibinfo{volume}{489} (\bibinfo{year}{2018}),
  \bibinfo{pages}{141--148}.
\newblock


\bibitem[Tang et~al\mbox{.}(2021)]%
        {tang2021cutqc}
\bibfield{author}{\bibinfo{person}{Wei Tang}, \bibinfo{person}{Teague Tomesh},
  \bibinfo{person}{Martin Suchara}, \bibinfo{person}{Jeffrey Larson}, {and}
  \bibinfo{person}{Margaret Martonosi}.} \bibinfo{year}{2021}\natexlab{}.
\newblock \showarticletitle{CutQC: using small quantum computers for large
  quantum circuit evaluations}. In \bibinfo{booktitle}{\emph{Proceedings of the
  26th ACM International Conference on Architectural Support for Programming
  Languages and Operating Systems}}. \bibinfo{pages}{473--486}.
\newblock


\bibitem[Tannu et~al\mbox{.}(2022)]%
        {tannu2022hammer}
\bibfield{author}{\bibinfo{person}{Swamit~S Tannu}, \bibinfo{person}{Poulami
  Das}, \bibinfo{person}{Ramin Ayanzadeh}, {and} \bibinfo{person}{Moinuddin~K
  Qureshi}.} \bibinfo{year}{2022}\natexlab{}.
\newblock \showarticletitle{HAMMER: Boosting Fidelity of Noisy Quantum Circuits
  by Exploiting Hamming Behavior of Erroneous Outcomes}. In
  \bibinfo{booktitle}{\emph{Proceedings of the Twenty-Seventh International
  Conference on Architectural Support for Programming Languages and Operating
  Systems}}. \bibinfo{pages}{529--540}.
\newblock


\bibitem[Tannu and Qureshi(2019)]%
        {tannu2019not}
\bibfield{author}{\bibinfo{person}{Swamit~S Tannu} {and}
  \bibinfo{person}{Moinuddin~K Qureshi}.} \bibinfo{year}{2019}\natexlab{}.
\newblock \showarticletitle{Not all qubits are created equal: a case for
  variability-aware policies for NISQ-era quantum computers}. In
  \bibinfo{booktitle}{\emph{Proceedings of the Twenty-Fourth International
  Conference on Architectural Support for Programming Languages and Operating
  Systems}}. \bibinfo{pages}{987--999}.
\newblock


\bibitem[Vikst{\aa}l et~al\mbox{.}(2020)]%
        {vikstaal2020applying}
\bibfield{author}{\bibinfo{person}{Pontus Vikst{\aa}l},
  \bibinfo{person}{Mattias Gr{\"o}nkvist}, \bibinfo{person}{Marika Svensson},
  \bibinfo{person}{Martin Andersson}, \bibinfo{person}{G{\"o}ran Johansson},
  {and} \bibinfo{person}{Giulia Ferrini}.} \bibinfo{year}{2020}\natexlab{}.
\newblock \showarticletitle{Applying the quantum approximate optimization
  algorithm to the tail-assignment problem}.
\newblock \bibinfo{journal}{\emph{Physical Review Applied}}
  \bibinfo{volume}{14}, \bibinfo{number}{3} (\bibinfo{year}{2020}),
  \bibinfo{pages}{034009}.
\newblock


\bibitem[Villalonga et~al\mbox{.}(2020)]%
        {villalonga2020establishing}
\bibfield{author}{\bibinfo{person}{Benjamin Villalonga},
  \bibinfo{person}{Dmitry Lyakh}, \bibinfo{person}{Sergio Boixo},
  \bibinfo{person}{Hartmut Neven}, \bibinfo{person}{Travis~S Humble},
  \bibinfo{person}{Rupak Biswas}, \bibinfo{person}{Eleanor~G Rieffel},
  \bibinfo{person}{Alan Ho}, {and} \bibinfo{person}{Salvatore Mandr{\`a}}.}
  \bibinfo{year}{2020}\natexlab{}.
\newblock \showarticletitle{Establishing the quantum supremacy frontier with a
  281 pflop/s simulation}.
\newblock \bibinfo{journal}{\emph{Quantum Science and Technology}}
  \bibinfo{volume}{5}, \bibinfo{number}{3} (\bibinfo{year}{2020}),
  \bibinfo{pages}{034003}.
\newblock


\bibitem[Wang(2019)]%
        {wang2019complex}
\bibfield{author}{\bibinfo{person}{Hengbin Wang}.}
  \bibinfo{year}{2019}\natexlab{}.
\newblock \emph{\bibinfo{title}{Complex Web-API Network Construction Based on
  Barabasi-Albert Model and Popularity-similarity Optimization Model}}.
\newblock \bibinfo{thesistype}{Ph.\,D. Dissertation}. \bibinfo{school}{Auckland
  University of Technology}.
\newblock


\bibitem[Wu et~al\mbox{.}(2021)]%
        {wu2021fractal}
\bibfield{author}{\bibinfo{person}{Xu Wu}, \bibinfo{person}{Linlin Zhang},
  \bibinfo{person}{Jia Li}, {and} \bibinfo{person}{Ruzhen Yan}.}
  \bibinfo{year}{2021}\natexlab{}.
\newblock \showarticletitle{Fractal statistical measure and portfolio model
  optimization under power-law distribution}.
\newblock \bibinfo{journal}{\emph{The North American Journal of Economics and
  Finance}}  \bibinfo{volume}{58} (\bibinfo{year}{2021}),
  \bibinfo{pages}{101496}.
\newblock


\bibitem[Xie et~al\mbox{.}(2022)]%
        {xie2022suppressing}
\bibfield{author}{\bibinfo{person}{Lei Xie}, \bibinfo{person}{Jidong Zhai},
  \bibinfo{person}{ZhenXing Zhang}, \bibinfo{person}{Jonathan Allcock},
  \bibinfo{person}{Shengyu Zhang}, {and} \bibinfo{person}{Yi-Cong Zheng}.}
  \bibinfo{year}{2022}\natexlab{}.
\newblock \showarticletitle{Suppressing ZZ crosstalk of Quantum computers
  through pulse and scheduling co-optimization}. In
  \bibinfo{booktitle}{\emph{Proceedings of the 27th ACM International
  Conference on Architectural Support for Programming Languages and Operating
  Systems}}. \bibinfo{pages}{499--513}.
\newblock


\bibitem[Zadorozhnyi and Yudin(2012)]%
        {zadorozhnyi2012structural}
\bibfield{author}{\bibinfo{person}{Vladimir~Nikolaevich Zadorozhnyi} {and}
  \bibinfo{person}{Evgenii~Borisovich Yudin}.} \bibinfo{year}{2012}\natexlab{}.
\newblock \showarticletitle{Structural properties of the scale-free
  Barabasi-Albert graph}.
\newblock \bibinfo{journal}{\emph{Automation and Remote Control}}
  \bibinfo{volume}{73}, \bibinfo{number}{4} (\bibinfo{year}{2012}),
  \bibinfo{pages}{702--716}.
\newblock


\bibitem[Zbinden et~al\mbox{.}(2020)]%
        {zbinden2020embedding}
\bibfield{author}{\bibinfo{person}{Stefanie Zbinden}, \bibinfo{person}{Andreas
  B{\"a}rtschi}, \bibinfo{person}{Hristo Djidjev}, {and}
  \bibinfo{person}{Stephan Eidenbenz}.} \bibinfo{year}{2020}\natexlab{}.
\newblock \showarticletitle{Embedding algorithms for quantum annealers with
  chimera and pegasus connection topologies}. In
  \bibinfo{booktitle}{\emph{International Conference on High Performance
  Computing}}. Springer, \bibinfo{pages}{187--206}.
\newblock


\end{thebibliography}


\end{document}